\documentclass{spie}

\usepackage{amsmath,amsfonts,amssymb}
\usepackage{graphicx}
\usepackage[colorlinks=true, allcolors=blue]{hyperref}
\usepackage{float}
\usepackage{array}
\newcolumntype{P}[1]{>{\centering\arraybackslash}p{#1}} 
\usepackage{xcolor}
\usepackage{gensymb}
\usepackage{textcomp}
\usepackage{soul}
\usepackage[perpage]{footmisc}
\title{Updated simulation tools for Roman coronagraph PSFs}

\author[a]{Kian Milani}
\affil[a]{James C. Wyant College of Optical Sciences, University of Arizona}
\author[b]{Ewan S. Douglas}
\affil[b]{Steward Observatory, University of Arizona \\
Massachusetts Institute of Technology}
\author[a]{Jaren Ashcraft}

\begin{document} 
\maketitle

\begin{abstract}
    The Nancy Grace Roman Space Telescope Coronagraph Instrument will be the first large scale coronagraph mission with active wavefront control to be operated in space and will demonstrate technologies essential to future missions to image Earth-like planets. Consisting of multiple coronagraph modes, the coronagraph is expected to characterize and image exoplanets at 1E-8 or better contrast levels. An object-oriented physical optics modeling tool called POPPY provides flexible and efficient simulations of high-contrast point spread functions (PSFs). As such, three coronagraph modes have been modeled in POPPY. In this paper, we present the recent testing results of the models and provide quantitative comparisons between results from POPPY and existing tools such as PROPER/FALCO. These comparisons include the computation times required for PSF calculations. In addition, we discuss the future implementation of the POPPY models for the POPPY front-end package WebbPSF, a widely used simulation tool for JWST PSFs. 
\end{abstract}

\section{Introduction}
The Roman Space Telescope will have two primary instruments on board, one of which is the Wide Field Instrument (WFI) designed for cosmology and dark energy research while the other is the coronagraph to be used for high-contrast imaging. Designed for the direct imaging of Jupiter sized planets along with debris disks (typical examples requiring ~1E-8 contrast), the coronagraph (also referred to as the Coronagraph Instrument, CGI) is integral for demonstrating the technologies required and best-suited for future missions such as HabEx\cite{habex} and LUVOIR\cite{luvoir} that will be interested in the more challenging effort of imaging Earth-like exoplanets near the habitable zone of nearby stars\cite{roman-info} (typical examples requiring ~1E-10 contrast). A primary focus of these technologies will be the CGI's demonstration of active wavefront control as it will be the first coronagraphic instrument in space to employ the hardware and software for sensing and control. 

\begin{figure}[H]
    \centering
    \includegraphics[scale=0.5]{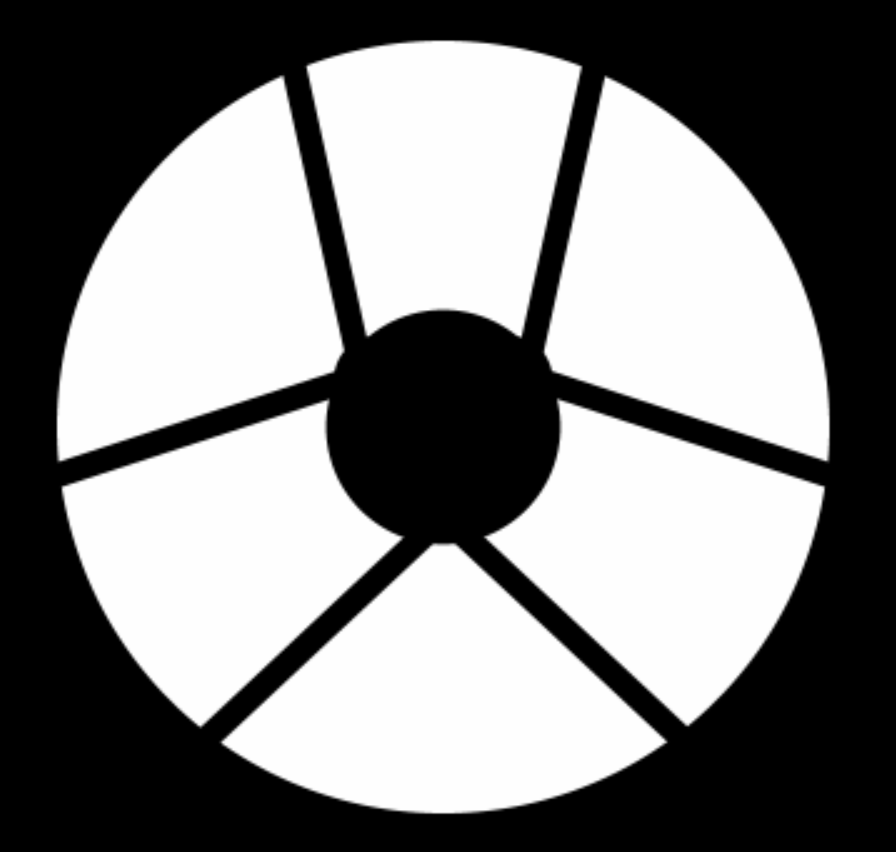}
    \caption{Illustration of the Roman pupil, which is $\sim$2.36m in diameter.}
    \label{fig:roman-pupil}
\end{figure}

Planned for launch in 2025 or later, Roman's development up to this point has been aided by the efforts put into modeling its performance using physical optics propagation (POP), which will continue to be an integral tool in Roman's further development. POP modeling is significant for coronagraphs because by suppressing on-axis starlight in order to achieve high contrast for off-axis sources such as exoplanets, the Point-Spread Functions (PSFs) evolve as a function of source position. Analyzing the PSFs is crucial to understanding the contrast ratio the CGI will be able to achieve. To obtain such PSFs for end-to-end models of the instrument, Fresnel propagation is used to propagate a complex electric field from optic to optic until the final image plane is reached. 

One of the major benefits of end-to-end modeling is the ability to include Optical Path Difference (OPD) errors and their mixing with amplitude errors (the Talbot effect) in the models of the system. Given the sensitivity of coronagraphic instruments, small OPD errors on the surfaces of optics can lead to very poor contrast ratios, even with no atmospheric turbulence affecting the wavefront. By measuring or estimating the OPDs on each optic of the system, they can be included in the POP models to analyze the affect on the PSFs. POP can also implement the affects of the deformable mirrors (DMs) so wavefront correction can also be simulated with specified DM maps.

The primary package utilized for simulations and modeling of the CGI has been PROPER\cite{krist_proper:_2007,krist_overview_2015,krist_wfirst_2017,krist_wfirst_2018}. PROPER is a package developed for POP available in IDL, MATLAB, and Python. It utilizes Fresnel and angular spectrum techniques to propagate a wavefront a given distance and includes other routines for modeling optical elements and applying their effects to a wavefront. This allows for a system to be modeled by propagating a wavefront from optic to optic to obtain a PSF. 

As of December 2019, the Phase-B models of the Roman CGI (previously known as WFIRST CGI) have been available through version 1.7 of the wfirst$\_$phaseb$\_$proper\footnote{\href{https://github.com/ajeldorado/proper-models/tree/master/wfirst_cgi/models_phaseb/python}{https://github.com/ajeldorado/proper-models/tree/master/wfirst$\_$cgi/models$\_$phaseb/python}} package. The Phase-B models are the ones that have been updated by reworking the models to operate with Physical Optic Propagation in Python (POPPY) rather than PROPER. As such, much of the data required for the models to operate, which was stored in FITS files meant to be used with wfirst$\_$phaseb$\_$proper, were altered for use in the updated simulation tools. The PROPER models are also what are used for the Fast Linearized Coronagraph Optimizer (FALCO) software. FALCO is an open-source software utilized for complex wavefront estimation and correction with routines for Electric Field Conjugation (EFC) and pair-wise probing\cite{falco}, two techniques utilized to obtain the previously mentioned DM maps that correct wavefront errors. FALCO uses PROPER in order to calculate the corrected wavefronts and for Roman, the models used in FALCO are the same as discussed here. 

\section{CGI Design and Specifications}
The CGI is designed for operation in a variety of modes. The basic collimating and refocusing optics for each mode will be the same, but the masks and filters used for different modes will be swapped when switching to a different mode of operation. Table \ref{tab:modes} lists the various modes along with their wavebands and intended field of view specifications. Note that only three of these modes will be tested on the ground, with the mode not to be tested being the SPC660 mode. The SPC660 mode has not been tested with the updated models, so only results of the other three modes will be discussed.

\begin{table}[H]
    \centering
    \caption{List of CGI modes including the Hybrid-Lyot Coronagraph (HLC) and the various Shaped-Pupil Coronagraphs (SPCs). The HLC and the SPC825 modes are designed for imaging and polarimetry while the other two SPC modes are designed for spectroscopy\cite{roman-info}.}
    \begin{tabular}{|P{2cm}|P{3cm}|P{2.5cm}|P{2.5cm}|P{2.5cm}|} \hline
    Mode & Center Wavelength [nm] & FWHM of Waveband [nm] & IWA-OWA [$\lambda/D$] & FOV Angular Coverage [$\degree$]\\ \hline 
    HLC575 & 575 & 70 & 3-9 & 360 \\ \hline 
    SPC660 & 660 & 112 & 3-9 & 2*65 \\ \hline 
    SPC730 & 730 & 122 & 3-9 & 2*65 \\ \hline 
    SPC825 & 825 & 94 & 5.4-20 & 360 \\ \hline 
    \end{tabular}
    \label{tab:modes}
\end{table}

\begin{figure}[H]
    \centering
    \includegraphics[scale=0.5]{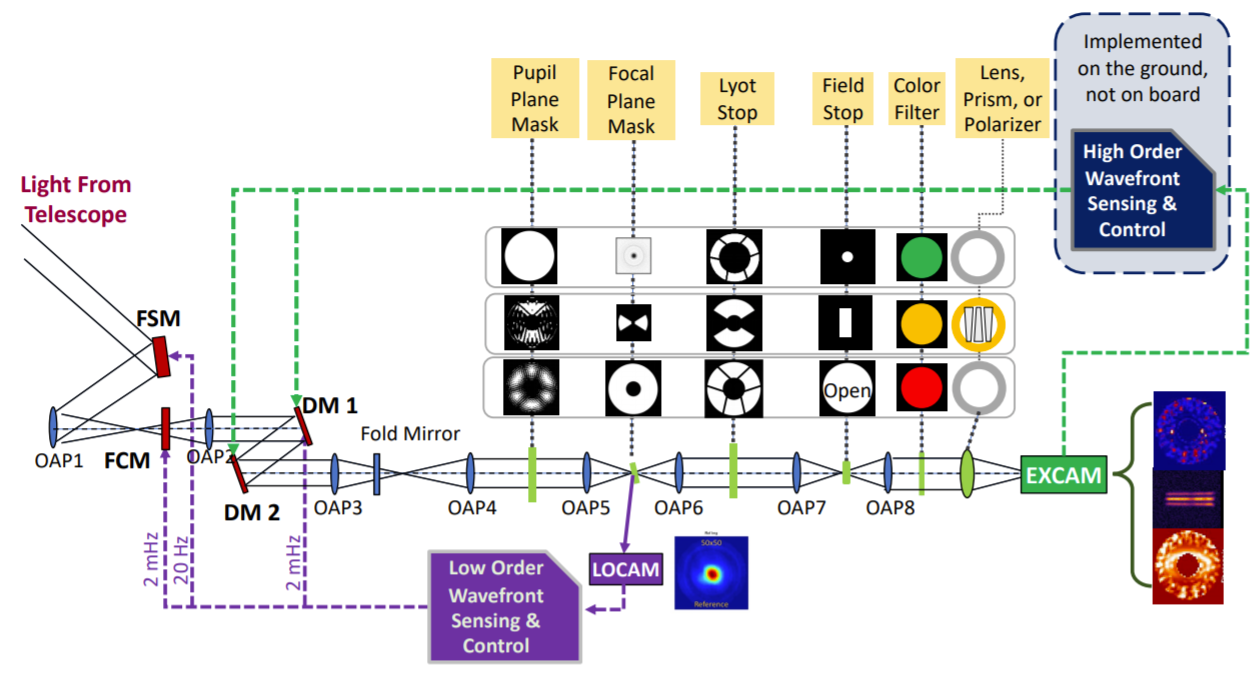}
    \caption{Roman coronagraph optical design as of December 2020 depicting the optics for each mode, reproduced from Kasdin et al\cite{roman-info}. The mode specific optics are illustrated at the according planes. The HLC contains no mask at the pupil-mask plane and utilizes a complex occulter for the FPM whereas the SPC modes utilize a binary pupil-mask and binary FPM designed for the intended FOV of the mode.}
    \label{fig:cgi-design}
\end{figure}

Figure \ref{fig:cgi-design} gives a general layout of the Roman CGI along with displaying the mode specific masks. These masks are the pupil plane masks (also known as shaped pupil masks or apodizers), focal plane mask (FPM), and Lyot stop. Also, in the PROPER models, there are various lens configurations after the filter, but for the purposes of POP testing and analyses, the basic focusing lens configuration is implemented. 

\section{Building each Mode's POPPY model}
In order to make the software user-friendly, POPPY operates in an object-oriented manner. It is similar to PROPER in that both utilize a wavefront type object with parameters that dictate the propagation of the wavefront. In POPPY, a wavefront object for Fresnel propagation is called a FresnelWavefront. Where POPPY and PROPER differ is that POPPY defines each optic in the system as an object/instance of a specific class type. Table \ref{tab:optics} in the appendix, gives the basic information necessary for each optic in the CGI, including what class each optic is defined as for POPPY. Each optic can then be sequentially added to an instance of a FresnelOpticalSystem class, where the distance from the previous optic is also given so POPPY propagates the wavefront to the correct plane. A FresnelOpticalSystem must first be initialized and given certain parameters such as the pupil diameter of the system, the number pixels spanning the pupil diameter, and the ratio of the beam size which determines the oversampling at pupil planes of the system. Note that the oversampling of a wavefront also determines the resolution at focal planes, where the more oversampling used, the higher the resolution at the focal plane. 

\begin{figure}
    \centering
    \includegraphics[scale=0.75]{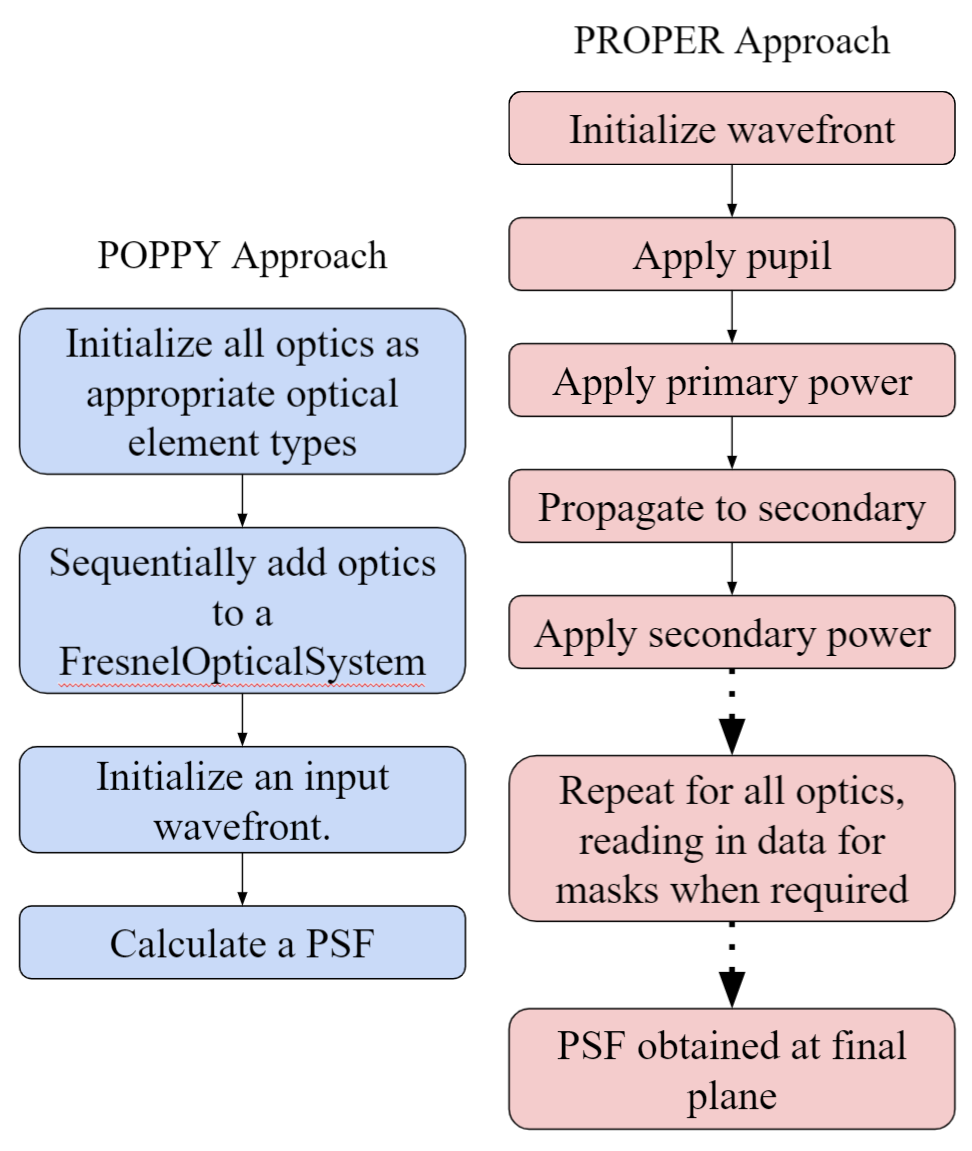}
    \caption{The two approaches between POPPY and PROPER for physical optics modeling of a given system. Note that the input wavefront is not required for POPPY, but was utilized for these results. POPPY performs the Fresnel propagation algorithm in the back-end of the software, so mathematically, the two softwares are very similar.}
    \label{fig:poppy-vs-proper}
\end{figure}

To calculate a PSF of a FresnelOpticalSystem, a simple routine is called, where POPPY uses the back-end Fresnel propagation algorithm to propagate the wavefront from optic to optic until the final image plane is reached. This makes POPPY slightly different from PROPER in that PROPER uses certain criteria to choose whether the wave is propagated using the Fresnel method or via the angular spectrum method. In addition to this difference, POPPY also uses a different convention for its phase and OPDs as it is designed to match optical engineering tools such as Zemax\footnote{\url{https://poppy-optics.readthedocs.io/en/latest/sign_conventions_for_coordinates_and_phase.html}}. The sign convention used is defined in Wyant and Creath\cite{basic-wf-theory}. This means the electric field arrays calculated at focal planes of the optical system using POPPY are rotated by $180\degree$ when compared to those from PROPER. In order for PROPER to calculate a PSF, a PROPER wavefront is first defined and the user propagates the wavefront given distances as well as applying optics to the wavefront through repeated use of the same routines. This is illustrated in Figure \ref{fig:poppy-vs-proper} along with the approach for POPPY. 

As for how each optic in the CGI models are initialized, many are defined as QuadraticLens objects as most of the optics are either collimating or focusing optics. In POPPY, a QuadraticLens object is used to add optical power to the phase of a wavefront in order to make it converge or diverge. The added phase is modeled purely as a quadratic surface that is determined by the focal length of the optic. Another commonly used optic type is the CircularAperture. This creates a simple circular aperture defined by a radius and this optic can be utilized to define the apertures of optics such as the powered or flat mirrors as well as being used for the field stop, which is specific to the HLC. However, the use of the circular apertures at various optics is not always necessary for accurate results as will be demonstrated in the results below. This is because the circular apertures of those optics often do not truncate the wavefront very much and so the diffraction affects from the edges of those optics are not critical for accuracy. 

Other optic types employed are the FITSOpticalElement and the FITSFPMElement. The FITSOpticalElements are primarily used to employ the pupil of the telescope along with pupil masks and DM patterns. POPPY does this by reading in both transmission and OPD data from given FITS files along with obtaining the pixelscales of the data such that the optic data can be applied to the wavefront at a given plane. However, the wavefront data will have a pixelscale calculated by POPPY through its propagation algorithm, so if there is a pixelscale mismatch between the wavefront and the optic data, POPPY attempts to interpolate the optic data to match the pixelscale to the wavefront and then apply the optic. This will be discussed more in depth when discussing the HLC as well as the models with individual optic OPDs as all OPDs are also initialized as FITSOpticalElements. 

The FITSFPMElement differs from a standard FITSOpticalElement in that is meant to be utilized only for focal plane masks. This optic type is a new addition to POPPY and was created for the purpose of replicating the PROPER results, although it can be used to model FPMs of other systems as well. Note that the code for this optic type is still pending final approval to be included in the most up to date version of POPPY\footnote{\url{https://github.com/spacetelescope/poppy}}, but for now, this feature can be found in a different POPPY repository\footnote{\url{https://github.com/kian1377/poppy}}. In the PROPER models, the FPM data for the two SPC modes did not include any pixelscales in units of meters/pixel, but only in units of ($\lambda/D$)/pixel. So the focal plane masks for those models could not be applied directly to the wavefront data because the pixelscales between the optic and the wavefront would not match and interpolation could not be done to force a match given the units were not equivalent. Instead, the PROPER models implemented an FFT and MFT (Matrix Fourier Transform) sequence in order to correctly apply the mask, so the FITSFPMElement is used in POPPY to do the same.

Another important factor in updating the simulation tools was including the effect of different polarization affects as these were included in the PROPER models and implemented by a custom routine utilizing data from the packages data files. In the PROPER models, the user would specify a polarization axis and before the wavefront propagation would begin, the transmission of the telescope pupil and the polarization aberrations would be applied to the wavefront. The specific polarization scenarios are described in the wfirst$\_$phaseb$\_$proper package, but they amount to different variations or combinations of input and output polarizations. When updating the simulations for POPPY, the feature allowing for a custom FresnelWavefront to be input to the system when calculating a PSF was used. To do so, a FresnelWavefront is first initialized with the corresponding parameters of pupil diameter and beam ratio to that of the FresnelOpticalSystem and then input into a slightly altered version of the wfirst$\_$phaseb$\_$proper routine that applies the polarization affects. The alterations in the routine are only made such that the routine can function with a FresnelWavefront from POPPY rather than a PROPER wavefront object.

\begin{figure}[H]
    \centering
    \includegraphics[scale=0.5]{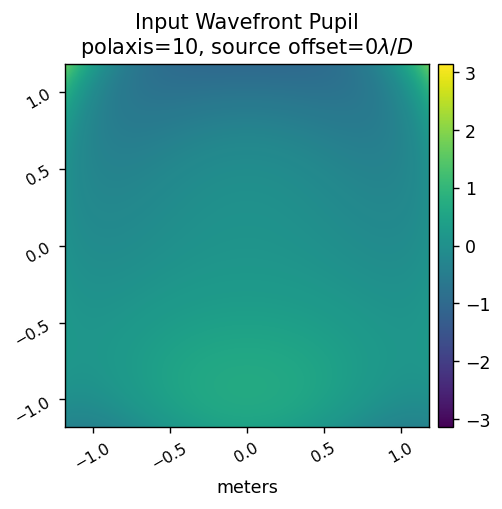}
    \includegraphics[scale=0.5]{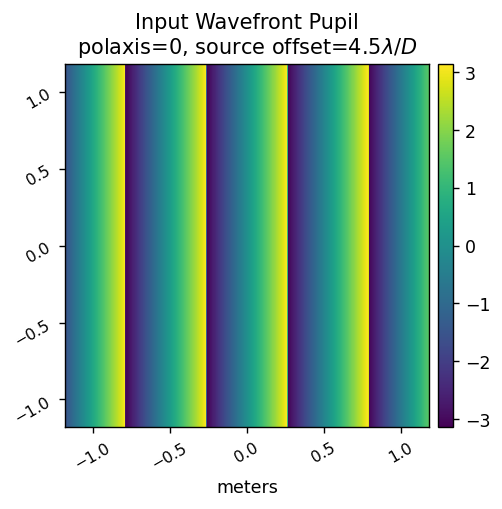}
    \includegraphics[scale=0.5]{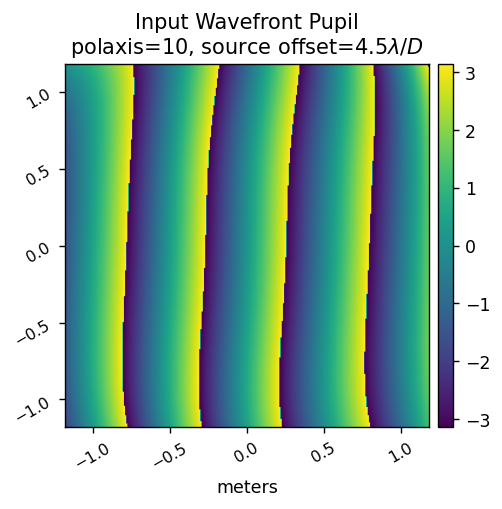}
    \caption{This figure demonstrates the phase of different input wavefronts. The left is the phase with the polarization axis parameter set to 10, corresponding to the scenario of $\pm45\degree$ input polarization with X and Y output polarization. The middle is the phase of a wavefront with a source offset of 4.5$\lambda/D$ in the x-direction, and the right is the combination of the left and middle.}
    \label{fig:inwave-variations}
\end{figure}

In addition to being utilized for the polarization affects, the input wavefront functionality of POPPY was also utilized to employ source offsets for off-axis PSFs. To do so, the user would provide source offset coordinates in units of $\lambda/D$ and the phase of the wavefront would be altered to reflect this offset. Figure \ref{fig:inwave-variations} displays the phase of a variety of input wavefronts. 

\subsection{POPPY HLC Mode}
Unlike the two SPC modes, the HLC does not utilize a shaped pupil mask and the FPM is a complex occulter affecting both the amplitude and phase of the wavefront. In addition, the HLC is designed to use the DMs to create a dark hole in the image plane whereas the two SPC modes create dark holes without the use of the DMs (assuming a perfect optical system).

In the wfirst$\_$phaseb$\_$proper model of the HLC, the Roman pupil diameter was set to 309pixels and the total wavefront array size was set to 1024pixels. However, because propagation from plane-to-plane is done by the user in PROPER, a user can change the total wavefront array size at a given plane. In the case of the HLC, the wavefront array size is changed to 2048pixels at OAP5 because this leads to a higher resolution at the FPM. This exact oversampling is key because in order to model the complex occulter in wfirst$\_$phaseb$\_$proper, 50 total FITS files were utilized. Of these files, 25 represented the real component of the occulter and 25 represented the imaginary component. The 25 pairs each correspond to a different wavelength and have differing pixelscales for each wavelength. This is because the pixelscale at the occulter plane is dependent on the oversampling of the wavefront array as well as the wavelength, so depending on what wavelength is being propagated, the closest matching wavelength of the occulter files is utilized. This ensures a close pixelscale match between the wavefront and the occulter data such that the occulter can be applied to the wavefront without any further interpolation being used to match pixelscales. Once the wavefront is propagated to the Lyot stop, the wavefront array size is changed back to 1024pixels, allowing for the propagation calculations to be completed more quickly. 

\begin{figure}[H]
    \centering
    \includegraphics[scale=0.5]{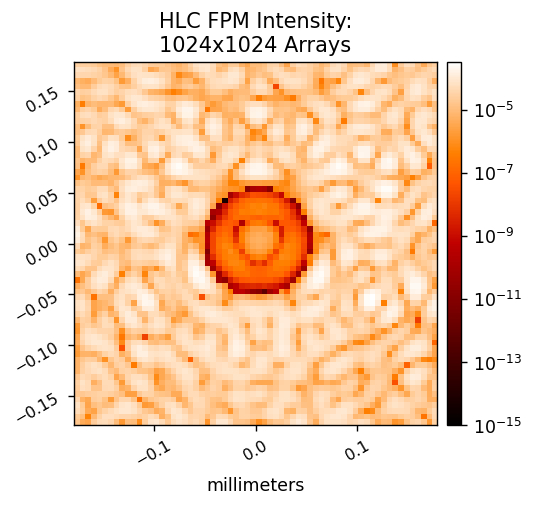}
    \includegraphics[scale=0.5]{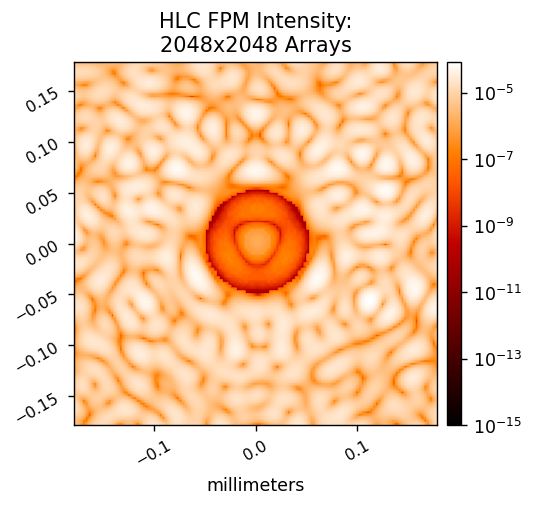}
    \caption{These are the intensities of the wavefronts calculated by POPPY at the FPM plane of the HLC after the complex occulter is applied. On the left is the result if the oversampling of the HLC mode is set to 1 while the right is the result when the oversampling is set to 2. The increased oversampling on the right makes the wavefront have a higher resolution at the focal plane. Specifically, the resolution is almost an exact match to the resolution of the complex occulter data files used for the HLC model. Note that the DM maps that create the dark-hole for the HLC are also used for these results.}
    \label{fig:hlc-fpm-comp}
\end{figure}

\begin{figure}[H]
    \centering
    \includegraphics[scale=0.4]{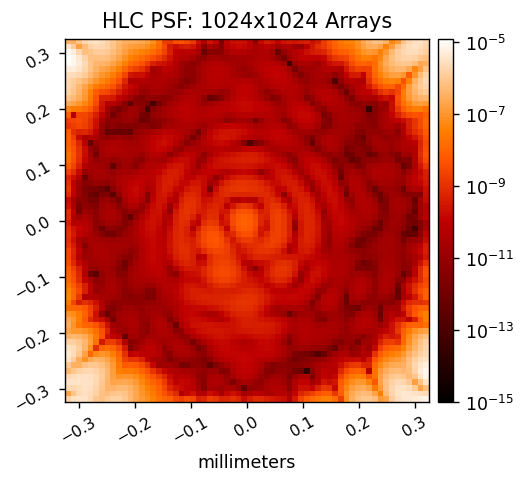}
    \includegraphics[scale=0.4]{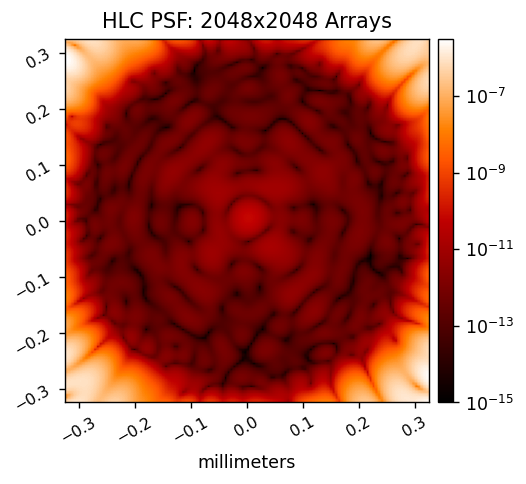}
    \caption{These are the PSF results calculated by POPPY when the oversampling is set to 1 (left) and when it is set to 2 (right). As a result of the lower resolution at the FPM, the results differ in that the PSF on the right is dimmer, indicating a better dark-hole. Results shown later on illustrate why the case on the right is more valid and why an oversampling of 2 was required for the HLC model.}
    \label{fig:hlc-psf-undersamp-comp}
\end{figure}

In order to recreate this in POPPY, the pupil diameter of the FresnelOpticalSystem is set to ~2.36m multiplied by 1024/309 with the number of pixels across the pupil being set to 1024. However, in order to acheive a close pixelscale match at the FPM, an oversampling factor of 2 is used for the FresnelOpticalSystem. This means 2048pixel wavefronts are propagated for the entire system. The accuracy of the wavefronts pixelscale to the complex occulter data is very important because the data was found to be very sensitive interpolation. Figure \ref{fig:hlc-fpm-comp} illustrates the difference between the wavefront of the HLC at the FPM plane when 2048pixel arrays are used versus 1024pixel arrays. The lower resolution in the wavefront and subsequently the occulter data leads to the difference between the two PSFs shown in Figure \ref{fig:hlc-psf-undersamp-comp}. Other than this, the same pupil and Lyot stop data are utilized for the POPPY model as the wfirst$\_$phaseb$\_$proper model. 

\subsection{POPPY SPC730 and SPC825 Modes}
For the SPC modes, the wfirst$\_$phaseb$\_$proper model would utilize 1000pixels across the pupil and the total wavefront array size was set to 2048pixels. The wavefront array size would then change to 4096pixels from the Lyot stop to the final image plane. For the POPPY models, the same value of 1000pixels was used across the pupil, but only 2x oversampling was used for the entire system, meaning 2000x2000pixel arrays were propagated. This significantly improves the speed at which the PSFs can be calculated. 

\begin{figure}[H]
    \centering
    \includegraphics[scale=0.5]{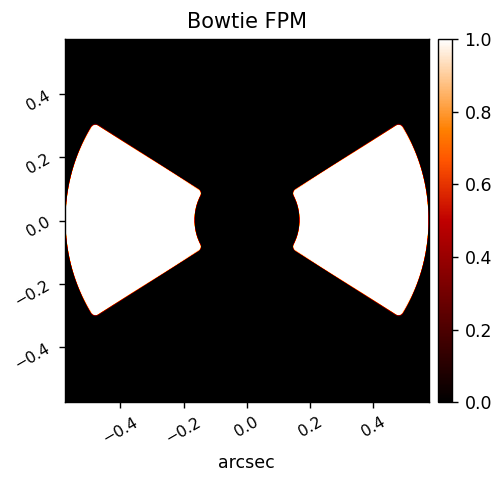}
    \includegraphics[scale=0.5]{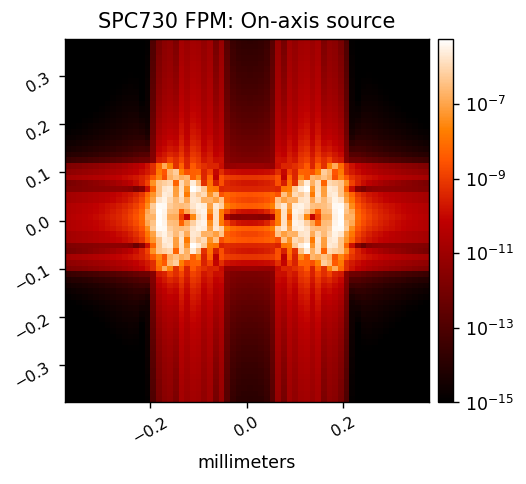}
    \includegraphics[scale=0.5]{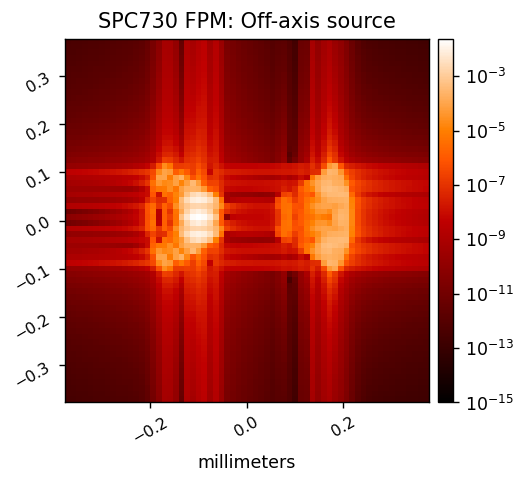}
    \caption{These images illustrate the bow tie shaped FPM for the SPC730 mode (left), the wavefront intensity after the FPM is applied for an on-axis source (middle), and the wavefront intensity after the FPM is applied for an off-axis source (right). The off-axis source is set to an angular offset of 4.5$\lambda/D$. The wavefronts were calculated by POPPY.}
    \label{fig:spcspec-fpm_comp}
\end{figure}

As for the configuration of the POPPY FresnelOpticalSystem, the most notable element is the FPM, which as mentioned before, is initialized with the FITSFPMElement class. This element requires known values of the FPM pixelscale, which was 0.05$\lambda/D$ for both SPC modes, the entrance pupil diameter, and the central wavelength of the mode. These parameters are required such that the MFT can be performed to the correct pixelscale and extent. The FFT and MFT sequence starts with an FFT to translate the wavefront to a virtual pupil plane. Then an inverse MFT is used to go back to the focal plane with a controlled pixelscale and dimension\cite{Soummer:07}. With the MFT matching the pixelscale of the wavefront to the FPM data, the FPM is applied to the wavefront. An MFT is then used to translate back to the virtual pupil and a final inverse FFT returns the wavefront to the focal plane, where it will have the same pixelscale as it began with. The standard POPPY propagation algorithm is then resumed. Figure \ref{fig:spcspec-fpm_comp} displays the results of applying the bow-tie shaped FPM for the SPC730 mode to the wavefront. 

\section{PSF Results of Each Mode}
For consistency, all the PSFs shown have array dimensions of 512x512 with a pixelscale of 0.1$\lambda/D$, where $\lambda$ is the central wavelength of the respective mode. It should be noted that the propagation algorithms do not generate the PSFs to the chosen pixelscale, but a PROPER routine is utilized to magnify the PSFs to the requested pixelscale\footnote{proper.prop$\_$magnify is the routine used for PSF interpolation.}. The extents of all the PSFs are generated from the respective pixelscale value in units of millimeters/pixel. The PSFs generated by the new POPPY models are all displayed alongside a PSF generated by wfirst$\_$phaseb$\_$proper with the equivalent settings (such as source offset, DMs, OPDs, etc...) employed. Also, as noted before, the original POPPY PSFs are rotated 180$\degree$ from the PROPER PSFs due to a sign difference in the respective propagation algorithms, so for visual comparison, POPPY PSFs are rotated by 180$\degree$ to match the orientation of PROPER PSFs. 

\subsection{HLC575 PSFs}
\label{sec:HLC575-PSFs}
\begin{figure}[H]
    \centering
    \includegraphics[scale=0.5]{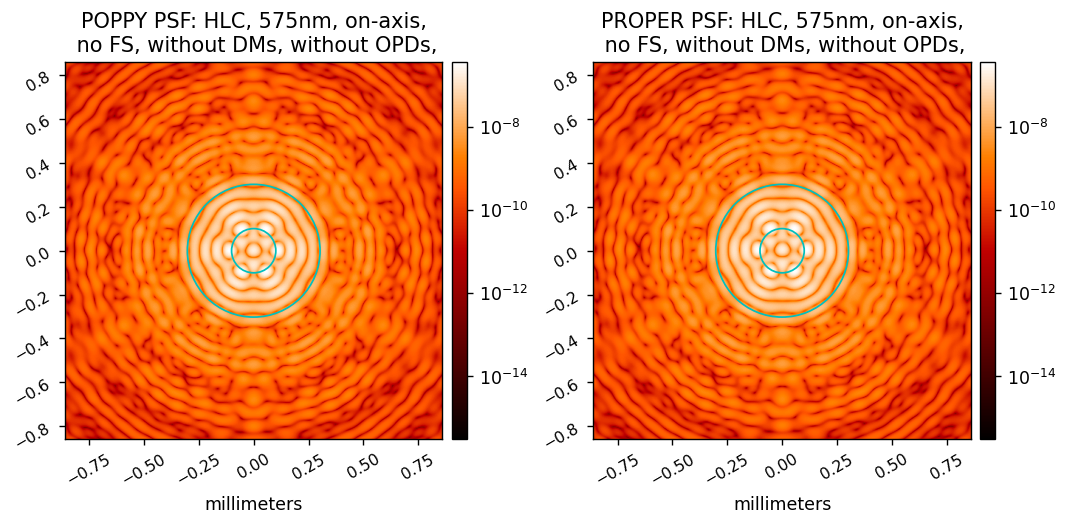}
    \caption{HLC PSF comparison with no field stop and no DMs utilized.}
    \label{fig:hlc-psfs-onax-nodms-nofs}
\end{figure}

\begin{figure}[H]
    \centering
    \includegraphics[scale=0.5]{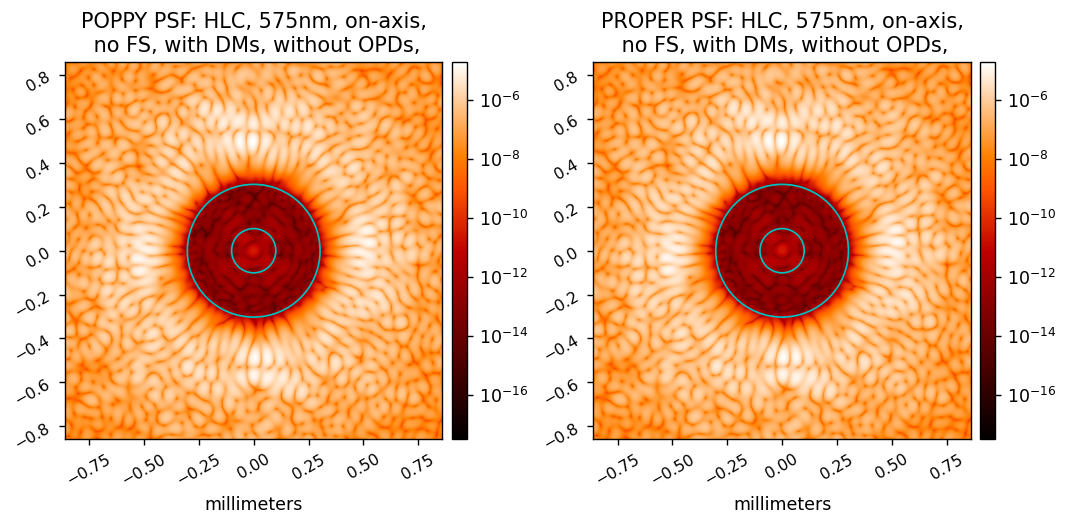}
    \caption{HLC PSF comparison with no field stop, but with DMs utilized.}
    \label{fig:hlc-psfs-onax-nofs}
\end{figure}

For the HLC PSFs, the 0.1$\lambda/D$ pixelscale corresponds to a value 3.361microns/pixel. This makes the total extent of each PSF range from -0.860mm to 0.860mm. Figure \ref{fig:hlc-psfs-onax-nodms-nofs} displays the PSF comparison with no DM corrections employed. The following pair shown in Figure \ref{fig:hlc-psfs-onax-nofs} illustrates the PSF comparisons with the DM maps included in the system. Together, these figures show how the HLC is designed to use the DMs to create a dark-hole, rather than using the DMs primarily for OPD correction.  

The following pairs of PSFs also employ the HLC field stop, which suppresses the excess light outside the outer working angle (OWA) of the HLC. While the structure in these PSFs appear very similar inside the OWA, the POPPY PSFs have much more ringing outside the OWA than the PROPER PSFs. This is a result of the diffraction from the edges of the field stop, although why exactly the PROPER PSFs do not have this same structure is unknown. 

\begin{figure}[H]
    \centering
    \includegraphics[scale=0.5]{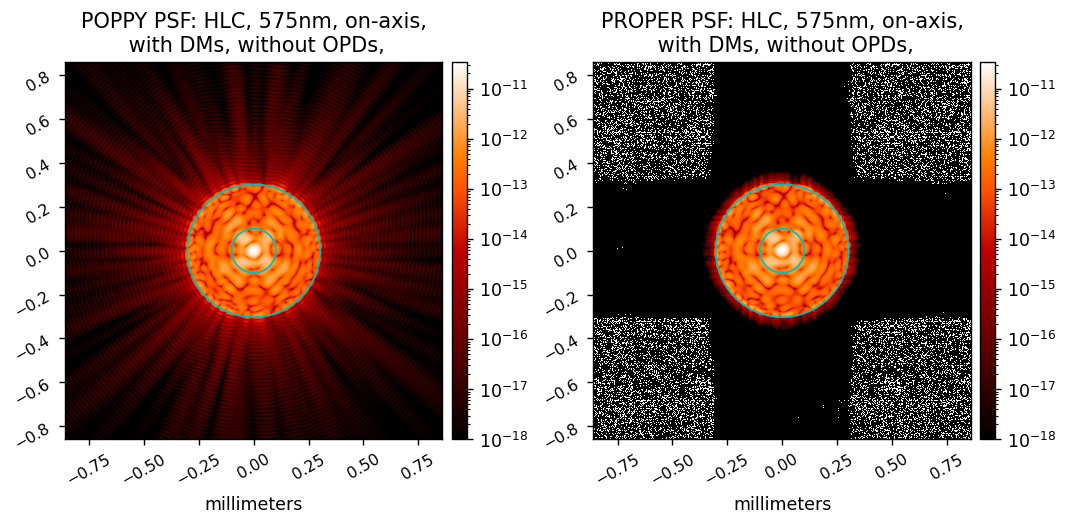}
    \caption{HLC PSF comparison with the field stop and DMs utilized.}
    \label{fig:hlc-psfs-onax-dms}
\end{figure}

\begin{figure}[H]
    \centering
    \includegraphics[scale=0.5]{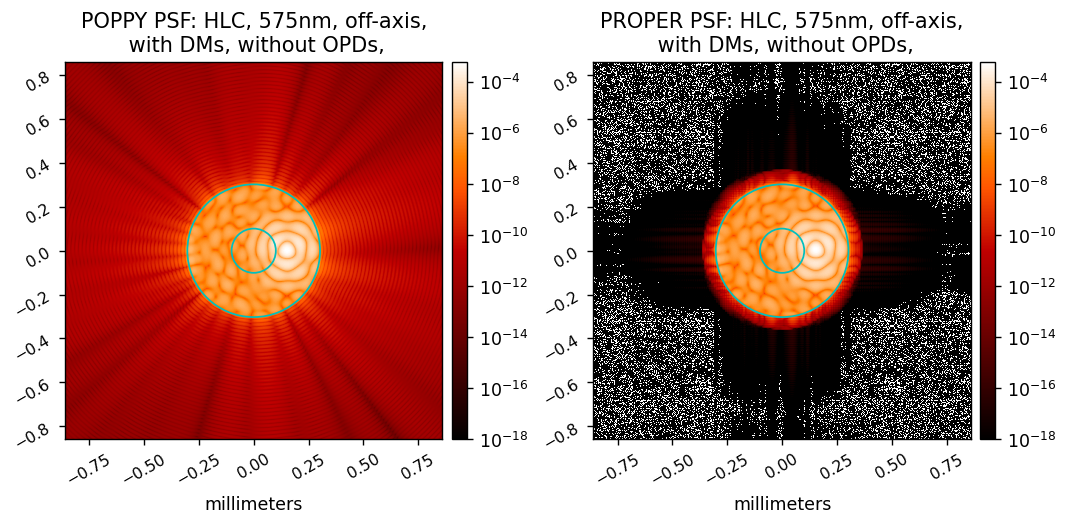}
    \caption{HLC PSF comparison with the field stop and DMs utilized for a source that is 4.5$\lambda/D$ off-axis.}
    \label{fig:hlc-psfs-offax-dms}
\end{figure}

\subsection{SPC730 PSFs}
\begin{figure}[H]
    \centering
    \includegraphics[scale=0.5]{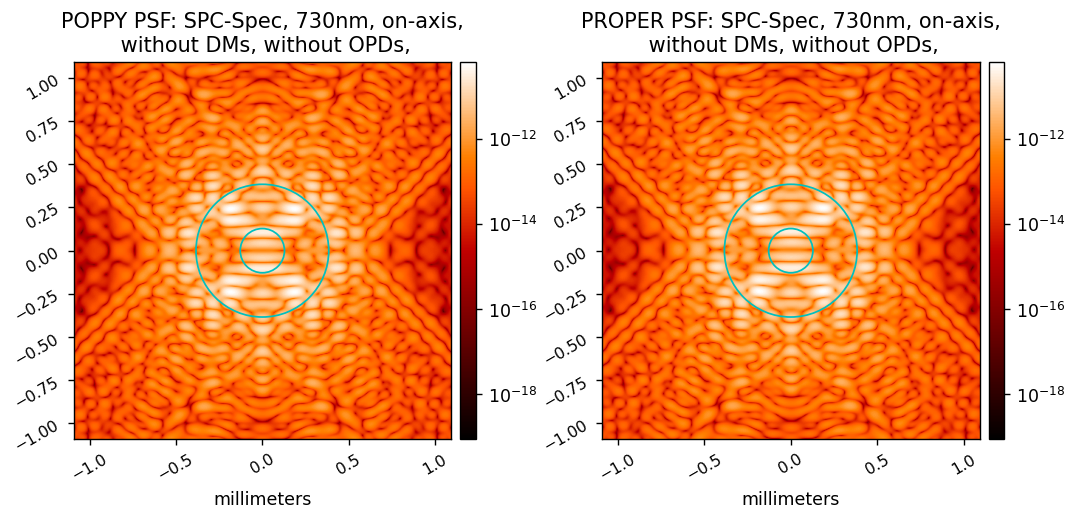}
    \caption{SPC730 PSF comparison without any DMs used as the SPC730 mode is not designed to use DMs to create the dark-hole unless OPDs are present.}
    \label{fig:spcspec-psfs-onax}
\end{figure}

For these PSFs, the 0.1$\lambda/D$ pixelscale corresponds to 4.267microns/pixel. This means the total extent of these PSFs ranges from -1.09mm to 1.09mm. Figure \ref{fig:spcspec-psfs-onax} shows the comparison between the POPPY and PROPER on-axis PSFs at the central wavelength of the mode while Figure \ref{fig:spcwide-psfs-onax-offlam} shows the results at a non-central wavelength. Both figures demonstrate agreement between POPPY and PROPER. 

\begin{figure}[H]
    \centering
    \includegraphics[scale=0.5]{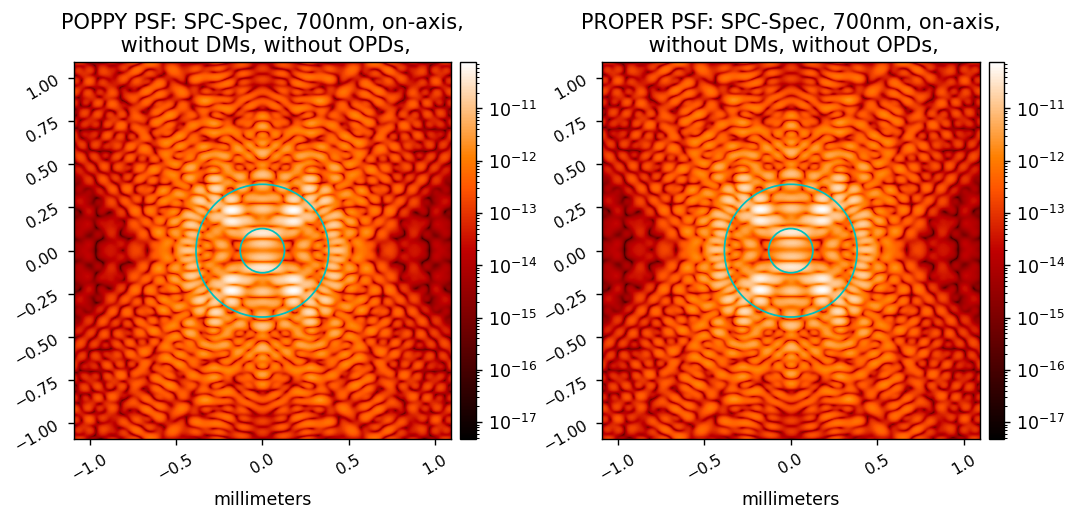}
    \caption{SPC730 PSF comparison for the wavelength of 700nm instead of the modes central wavelength.}
    \label{fig:spcspec-psfs-onax-offlam}
\end{figure}

\begin{figure}[H]
    \centering
    \includegraphics[scale=0.5]{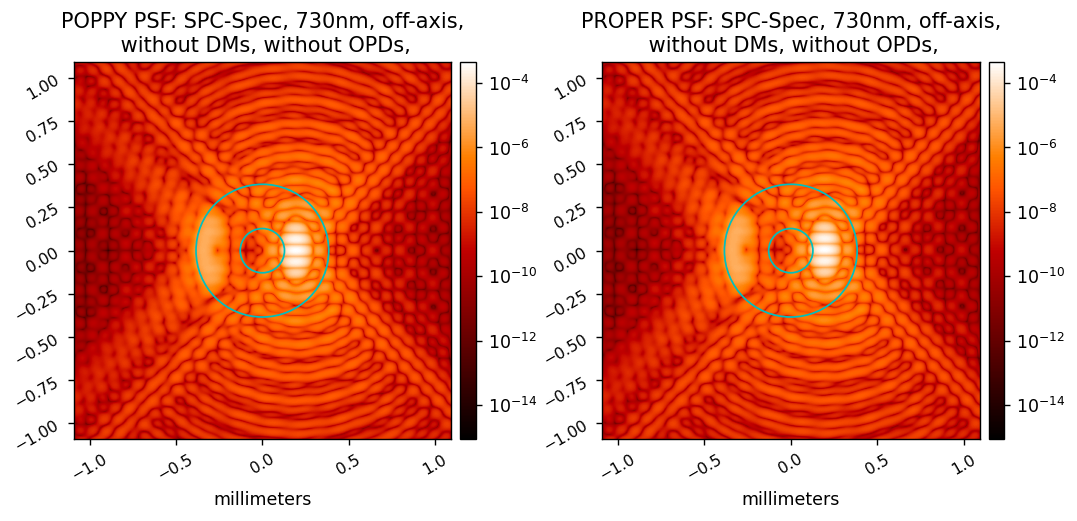}
    \caption{SPC730 PSF comparison for a source set to 4.5$\lambda/D$ off-axis.}
    \label{fig:spcspec-psfs-offax}
\end{figure}

Figure \ref{fig:spcspec-psfs-offax} shows the results for an off-axis case where POPPY and PROPER also demonstrate agreement.

Finally, the following PSFs are the results from POPPY and PROPER when individual optic apertures are also used. This is done by applying a circular aperture at the plane of each optical element. As mentioned earlier, the apertures do not make a large difference, as these PSFs are almost identical to those shown in Figure \ref{fig:spcspec-psfs-onax}. This is why the apertures are not used for most of the testing performed. 

\begin{figure}[H]
    \centering
    \includegraphics[scale=0.5]{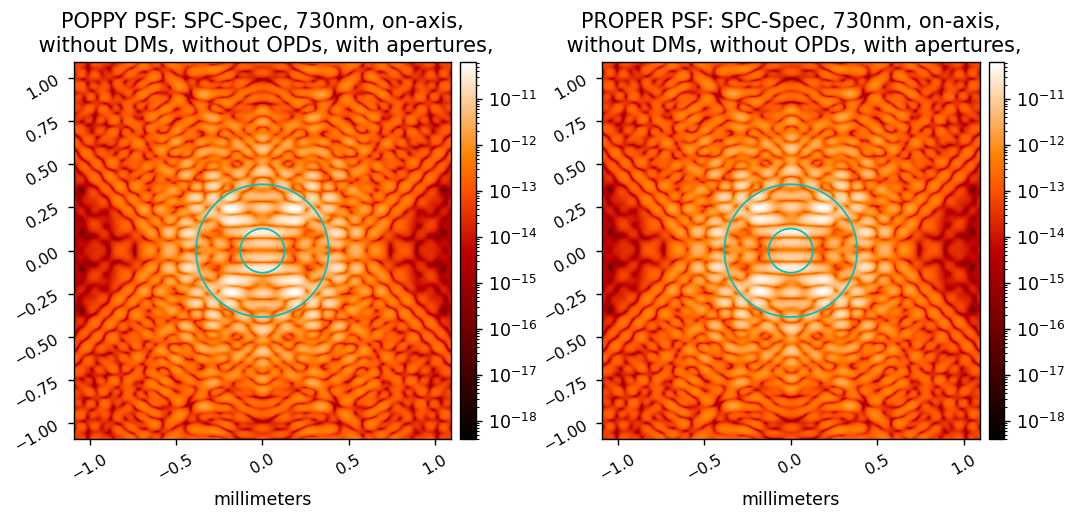}
    \caption{SPC730 PSF comparison when individual optic apertures are used. There is very little difference between these PSFs and those in Figure \ref{fig:spcspec-psfs-onax}, illustrating why apertures are often not required for physical optics models.}
    \label{fig:spcspec-psfs-onax-aps}
\end{figure}

\subsection{SPC825 PSFs}
\begin{figure}[H]
    \centering
    \includegraphics[scale=0.5]{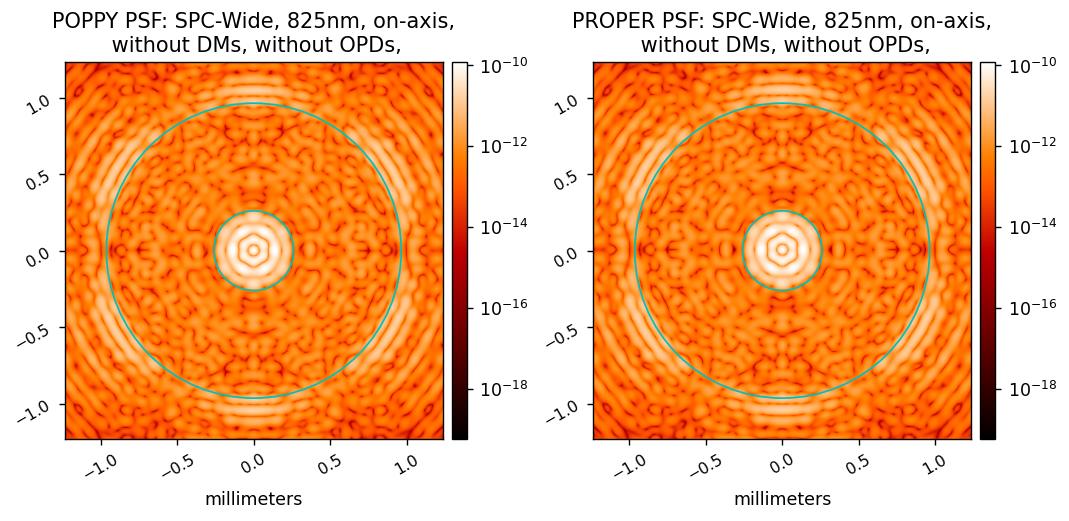}
    \caption{SPC825 PSF comparison for an on-axis source with no DMs or OPDs used.}
    \label{fig:spcwide-psfs-onax}
\end{figure}

\begin{figure}[H]
    \centering
    \includegraphics[scale=0.5]{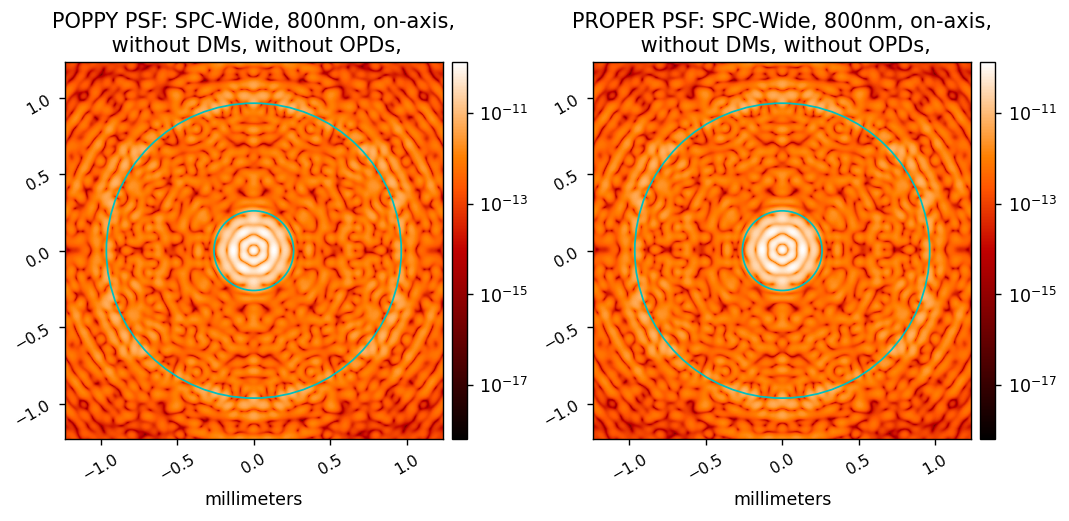}
    \caption{SPC825 PSF comparison for an on-axis source, but at a wavelength of 800nm rather than the modes central wavelength.}
    \label{fig:spcwide-psfs-onax-offlam}
\end{figure}

For this mode, the 0.1$\lambda/D$ pixelscale corresponds to 4.822microns/pixel, making the total extent of the PSFs range from -1.23mm to 1.23mm. Figures \ref{fig:spcwide-psfs-onax} and \ref{fig:spcwide-psfs-onax-offlam} illustrate the pairs of on-axis PSFs for both central and non-central wavelength cases. Like the SPC730 PSFs, these also demonstrate agreement between the models. 

\begin{figure}[H]
    \centering
    \includegraphics[scale=0.5]{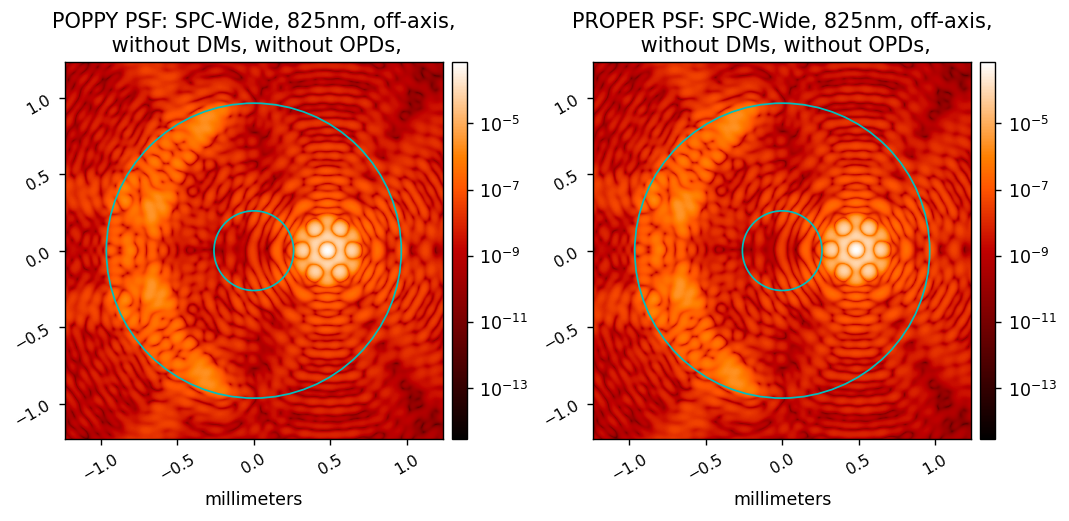}
    \caption{SPC825 PSF comparison for a source set to 10$\lambda/D$ off-axis. }
    \label{fig:spcwide-psfs-offax}
\end{figure}

Finally, Figure \ref{fig:spcwide-psfs-offax} show the case of off-axis PSFs in order to illustrate agreement between this case.

\section{PSFs with individual optic OPDs}
\begin{figure}[H]
    \centering
    \includegraphics[scale=0.5]{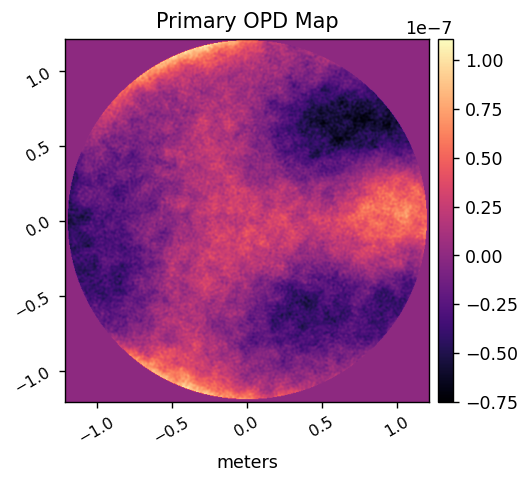}
    \caption{OPD map for Roman's primary mirror used in the PROPER models.}
    \label{fig:primary-opd-map}
\end{figure}

When adding OPDs to the POPPY models, the exact OPD maps included in the wfirst$\_$phaseb$\_$proper data files were not utilized. The PROPER models would use a PROPER routine\footnote{proper.prop$\_$errormap is the routine used for the optic OPD maps.}, which would use the OPD data of the given optic and match the pixelscale of the OPD data to that of the wavefront generated from propagation. The OPD map generated by the PROPER routine for every optic was saved as a FITS file that includes the pixelscale of the data so it could be implemented in POPPY as a FITSOpticalElement. There is an added benefit to this which is if the pupil diamater in units of pixels is the same as that used in the PROPER models, the pixelscales of those wavefronts in POPPY will be the same (or at least very close), so interpolation is no longer required and the total computation time is reduced. Figure \ref{fig:primary-opd-map} shows an example of one of the OPD maps used for the PROPER models. 

\begin{figure}[H]
    \centering
    \includegraphics[scale=0.5]{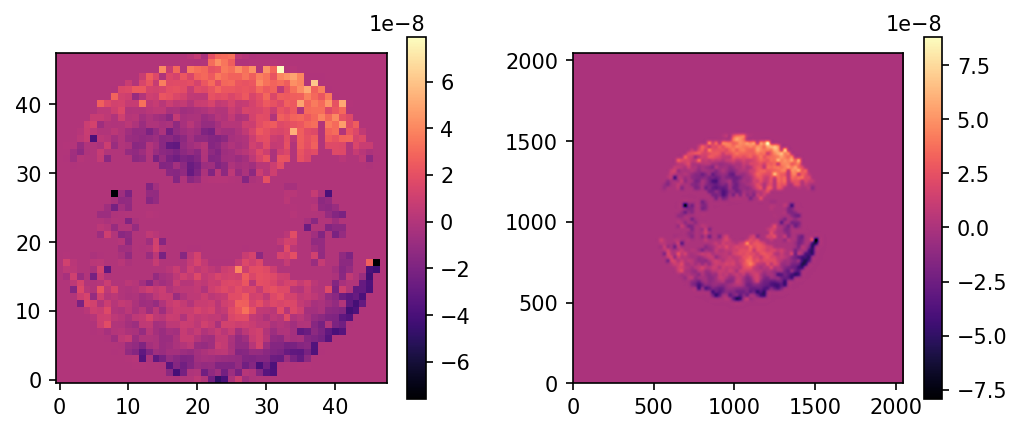}
    \caption{To the left is the original 48x48 grid for the DM's actuators and to the right is the OPD map of the actuators after the PROPER routine is used to apply the DM to the wavefront. The specific DM map shown is for the SPC730 mode's first DM.}
    \label{fig:dm-before-after}
\end{figure}

Also, in order for the modes to operate with DMs, the wfirst$\_$phaseb$\_$proper package included example DM maps for each mode. These DM maps were saved as 48x48 data representing the actuator positioning of the 48x48 DMs where the actuator spacing was 0.9906mm. PROPER would implement the DMs through another PROPER routine\footnote{proper.prop$\_$dm is the routine used for the DM actuator grids.} that would include the affect of an actuator influence function and generate a DM map with the same pixelscale of the wavefront at the DM plane, allowing the OPD map for the DM actuators to be directly applied to the wavefront. For convenience, when creating the POPPY models, the example DM maps were not utilized, but the results of the DM maps from the PROPER routine were saved as new FITS files that include the pixelscale. Each DM is then implemented into the FresnelOpticalSystem as a FITSOpticalElement. Figure \ref{fig:dm-before-after} shows the original 48x48 grid for the DMs actuators alongside the OPD map of the DM actuators after the PROPER routine is utilized to apply the DM to the wavefront. 

Note that all the PSFs for the modes with OPDs utilized have the same respective pixelscales as without the OPDs. Also, all results shown with OPDs used include polarization aberrations for the case of the polarization axis parameter being set to 10. 

\subsection{HLC575 PSFs with OPDs}
\begin{figure}[H]
    \centering
    \includegraphics[scale=0.5]{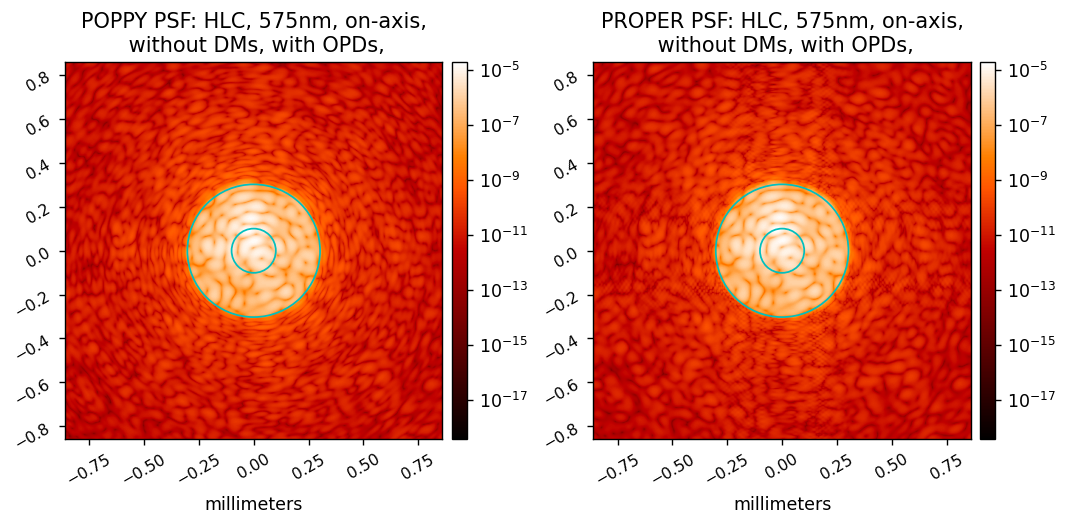}
    \caption{HLC575 PSF comparison with the OPDs applied and no DMs used for correction. because of the lack of correction, there is no dark-hole inside the OWA.}
    \label{fig:hlc-psfs-onax-opds}
\end{figure}

\begin{figure}[H]
    \centering
    \includegraphics[scale=0.5]{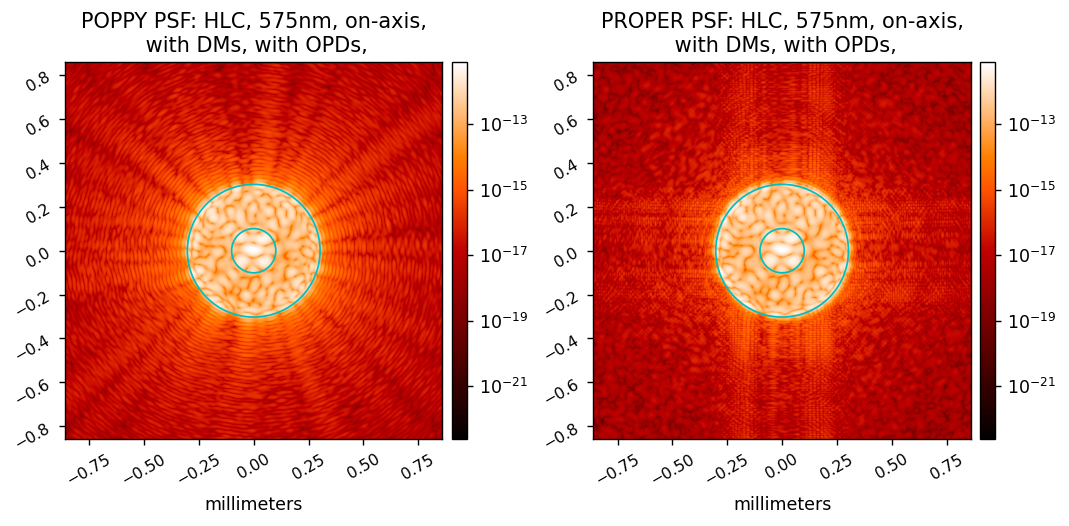}
    \caption{HLC575 PSF comparison with OPDs and DMs for correction. Both models illustrate a well-corrected dark-hole.}
    \label{fig:hlc-psfs-onax-opds-dms}
\end{figure}

The first pair of PSFs shown in Figure \ref{fig:hlc-psfs-onax-opds} is for the case of the HLC with no DM maps used. This is why there is not a corrected dark-hole region. The second pair of PSFs in Figure \ref{fig:hlc-psfs-onax-opds-dms} includes the DM settings for the HLC with OPDs, which is why there is a much dimmer dark-hole. 

\begin{figure}[H]
    \centering
    \includegraphics[scale=0.5]{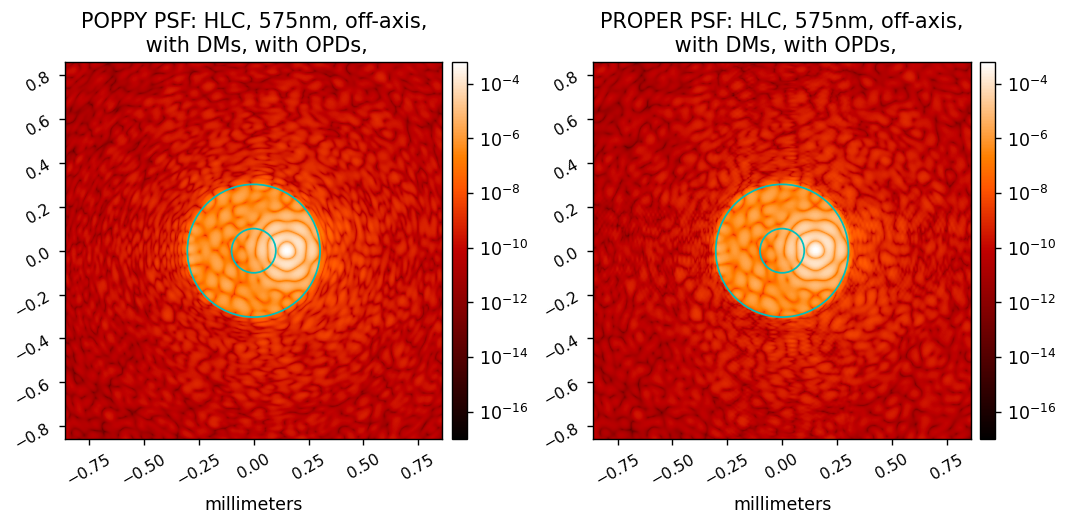}
    \caption{HLC575 PSF comparison with OPDs and DMs for an off-axis source. When compared with the PSFs from Figure \ref{fig:hlc-psfs-onax-opds-dms}, it is clear there is a high-contrast between the on-axis source and off-axis source.}
    \label{fig:hlc-psfs-offax-opds-dms}
\end{figure}

Figure \ref{fig:hlc-psfs-offax-opds-dms} shows the PSFs with the same settings as for the previous figure, but for an off-axis source. All the PSFs in this section also use the HLC field stop and overall, PROPER and POPPY demonstrate agreement inside the OWA.

\subsection{SPC730 PSFs with OPDs}
\begin{figure}[H]
    \centering
    \includegraphics[scale=0.5]{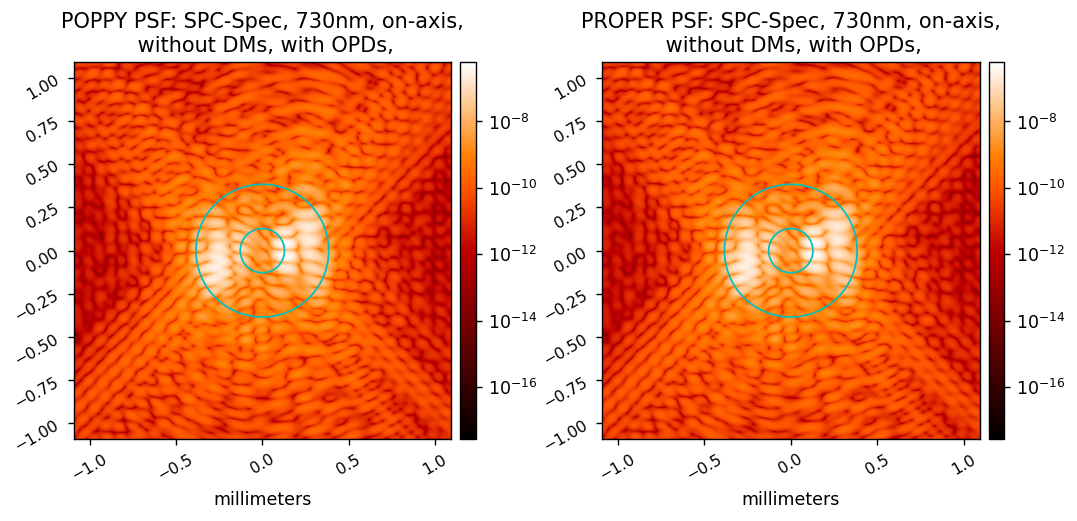}
    \caption{SPC730 PSF comparison when OPDs are applied with no DMs used for correction. Both dark-hole regions that previously made up a bow tie shape are far brighter due to the lack of aberration correction.}
    \label{fig:spcspec-psfs-onax-opds}
\end{figure}

\begin{figure}[H]
    \centering
    \includegraphics[scale=0.5]{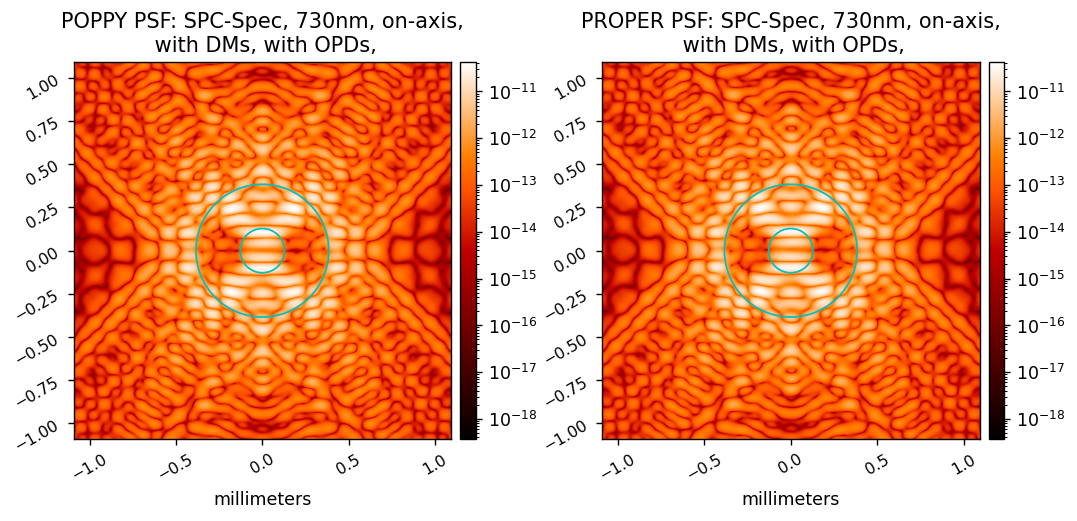}
    \caption{SPC730 PSF comparison with OPDs and DMs used. The bow tie shaped dark-hole is restored.}
    \label{fig:spcspec-psfs-onax-opds-dms}
\end{figure}

Like the HLC, the OPD aberrations for the SPC730 mode creates a very bright dark-hole. When the DM maps are also used though, the bow-tie shaped dark-hole is restored such that the instrument can meet the high-contrast requirements. This is demonstrated in Figures \ref{fig:spcspec-psfs-onax-opds} and \ref{fig:spcspec-psfs-onax-opds-dms} respectively. 

The next PSFs are for the off-axis case in order to illustrate the high-contrast the instrument can acheive. 

\begin{figure}[H]
    \centering
    \includegraphics[scale=0.5]{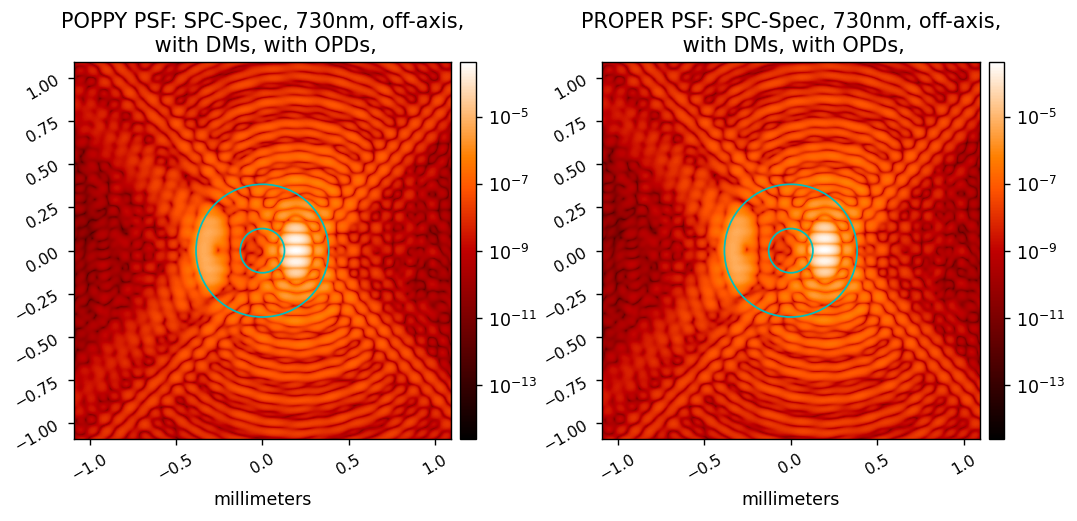}
    \caption{SPC730 PSF comparison with OPDs and DMs used for an off-axis source.}
    \label{fig:spcspec-psfs-offax-opds-dms}
\end{figure}

\subsection{SPC825 PSFs with OPDs}
\begin{figure}[H]
    \centering
    \includegraphics[scale=0.5]{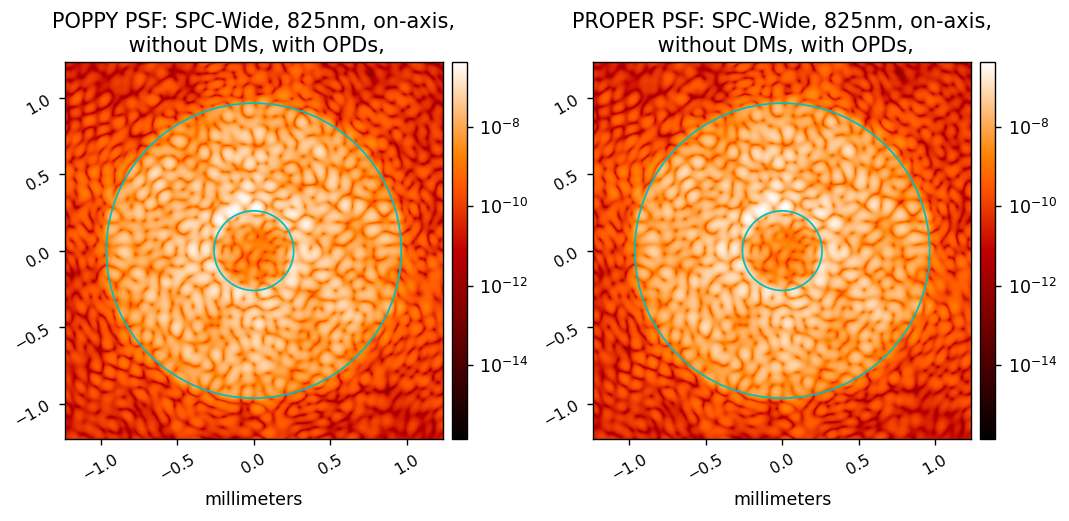}
    \caption{SPC825 PSF comparison with OPDs applied, but no DMs used for correction. Again, POPPY and PROPER show very similar results with the aberrations causing the dark-hole region to be filled with excess light form the on-axis source.}
    \label{fig:spcwide-psfs-onax-opds}
\end{figure}

\begin{figure}[H]
    \centering
    \includegraphics[scale=0.5]{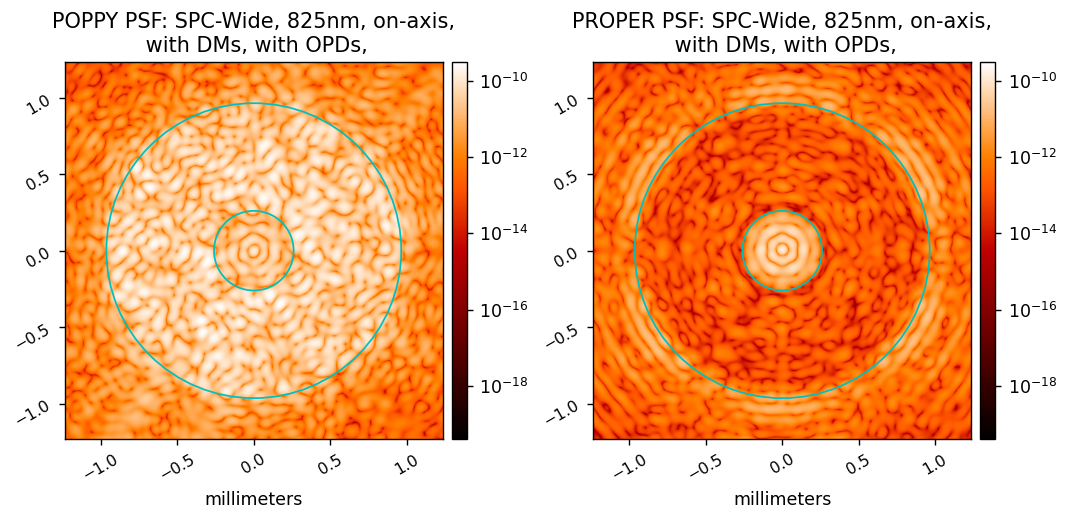}
    \caption{SPC825 PSF comparison with the OPDs and DM maps applied. Unlike the previous modes with OPDs, POPPY and PROPER come to different results, with PROPER showing a well-corrected dark-hole as expected. The exact reason for the POPPY PSF not having a corrected dark-hole is still being investigated, but this is the only case in which significant disagreement was found.}
    \label{fig:spcwide-psfs-onax-opds-dms}
\end{figure}

For this mode, Figure \ref{fig:spcwide-psfs-onax-opds} also shows the need for wavefront correction as there is no longer a dark-hole that can provide high-contrast. However, the following PSFs in Figure \ref{fig:spcwide-psfs-onax-opds-dms} is the only case found where there is significant disagreement between POPPY and PROPER. In those PSFs, the OPDs and DMs are utilized, and while there is a small region inside the IWA where POPPY and PROPER seem to agree, the region between the IWA and OWA is significantly different, with the PROPER PSF showing a well-corrected dark-hole while the POPPY PSF is still far brighter. 

\begin{figure}[H]
    \centering
    \includegraphics[scale=0.5]{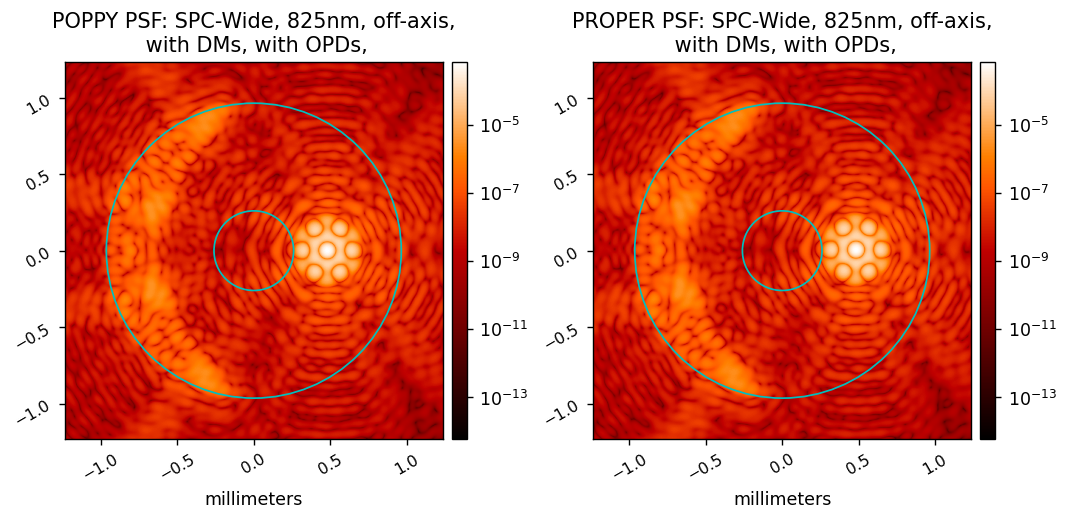}
    \caption{SPC825 PSF comparison with OPDs and DMs used for an off-axis source. While the previous figure had differing results, POPPY and PROPER agree for the off-axis source's PSF.}
    \label{fig:spcwide-psfs-offax-opds-dms}
\end{figure}

Further analysis showed that the on-axis POPPY PSF with OPDs and DMs is 3289x brighter between 5.4$\lambda/D$ and 20$\lambda/D$, or approximately 3 orders of magnitude brighter. The exact cause of this disagreement is still unknown and is still being investigated. Curiously, when the OPDs and DMs are used, the off-axis PSFs in Figure \ref{fig:spcwide-psfs-offax-opds-dms} do show agreement between POPPY and PROPER. 

\section{Computation Time Comparisons}
The computation time of a PSF is highly dependent on the size of the wavefront being propagated. As mentioned before, the wfirst$\_$phaseb$\_$proper models change the wavefront array size at certain planes in the system whereas the POPPY models use a uniform size throughout propagation. Therefore, the comparisons being made will not be 1-to-1, but nonetheless illustrate the difference in the updated simulation tools. All computation testing performed was done on a UArizona HPC Puma node. The specifications of the node are given in the table \ref{tab:hpc-specs}. 

\begin{table}[H]
    \centering
    \caption{Specifications of the UArizona HPC node utilized for all PSF computations.}
    \begin{tabular}{|P{2cm}|P{2cm}|P{1.6cm}|P{3.5cm}|P{3cm}|P{1.5cm}|} \hline
    UArizona HPC Node & Model & OS & CPU & GPU & Memory \\ \hline
    Puma 
    & Penguin Altus XE2242
    & CentOS 7
    & 2x AMD EPYC 7642 48-core (8-cores utilized), 2.4GHz
    & NVIDIA V100S (32GB GPU Memory)
    & 512GB \\ \hline
    \end{tabular}
    \label{tab:hpc-specs}
\end{table}

Both PROPER and POPPY also have the capability of implementing different packages that increase the speed of the FFTs used in propagation. The package common to both PROPER and POPPY is the pyFFTW package, which is a python wrapper for the C subroutine library FFTW and can be installed directly into the python environment to enable it. Other than pyFFTW, both PROPER and POPPY implement MKL FFTs, but with different methods. PROPER utilizes the Intel MKL package whereas POPPY utilizes the mkl$\_$fft package. Intel MKL is a package that must be installed locally whereas mkl$\_$fft can be installed directly into a python environment similar to pyFFTW. POPPY also has GPU based FFT methods involving the use of either CUDA, which is available through numba, or OpenCL, which is available with the PyOpenCL package. For this research, only comparisons with POPPY's PyOpenCL implementation were performed. 

The computation times listed in table \ref{tab:comp-times} are found by taking the average of 5 different PSF calculations. Note that the wavefront sizes being propagated for the POPPY models are 2048x2048 for the HLC and 2000x2000 for the two SPC modes. As mentioned before, the wavefront sizes in the PROPER models are changed at specific planes in the system, which is a major factor in why PROPER is faster when it comes to the HLC, but slower when it comes to the two SPC modes. 

\begin{table}[H]
    \centering
    \caption{Comparison of PSF computation times for the different coronagraph modes. All values are in units of seconds.}
    \begin{tabular}{|P{1.5cm}|P{1.5cm}|P{1.6cm}|P{1.5cm}|P{1.6cm}|P{1.5cm}|P{1.5cm}|} \hline
    Mode & PROPER & PROPER (pyFFTW) & POPPY & POPPY (pyFFTW) & POPPY (mkl$\_$fft) & POPPY (OpenCL) \\ \hline 
    HLC575 & 15.48 & 10.66 & 102.25 & 25.49 & 27.68 & 25.14 \\ \hline
    SPC730 & 172.47 & 96.87 & 28.23 & 25.25 & 26.00 & 25.11 \\ \hline
    SPC825 & 174.05 & 97.75 & 28.67 & 26.91 & 27.80 & 26.20 \\ \hline
    HLC575 (OPDs) & 21.72 & 16.30 & 117.04 & 32.15 & 32.39 & 29.19 \\ \hline
    SPC730 (OPDs) & 203.62 & 128.29 & 44.57 & 41.60 & 41.24 & 41.01 \\ \hline
    SPC825 (OPDs) & 192.78 & 116.94 & 45.32 & 42.76 & 42.35 & 41.59 \\ \hline
    \end{tabular}
    \label{tab:comp-times}
\end{table}

\section{WebbPSF Implementation}
WebbPSF is a widely used tool for modeling the James Webb Space Telescope uses POPPY as its propagation tool\cite{webbpsf}. Roman coronagraph models included in the latest package release (v0.9.1) are limited to an early version of the SPC mask designs without OPDs or DMs. The updated POPPY models described here are planned to be released through a future WebbPSF release to provide a familiar, user-friendly experience. The WebbPSF architecture allows the user to compare predefined operating modes.

Initial results of the new coronagraph masks when implemented in the WebbPSF framework are displayed below, where the pixelscales are also 0.1$\lambda/D$. However, the interpolation method used to adjust the pixelscales was a Scipy\footnote{scipy.ndimage.zoom was the interpolation method used for initial WebbPSF results} routine instead of the PROPER routine. The HLC PSFs using the SciPy interpolation method were found to be 1.0474x brighter on average than the POPPY PSFs that used the PROPER interpolation. Therefore, the SciPy interpolation results for the HLC are within 5$\%$ of the results previously shown.

\begin{figure}[H]
    \centering
    \includegraphics[scale=0.45]{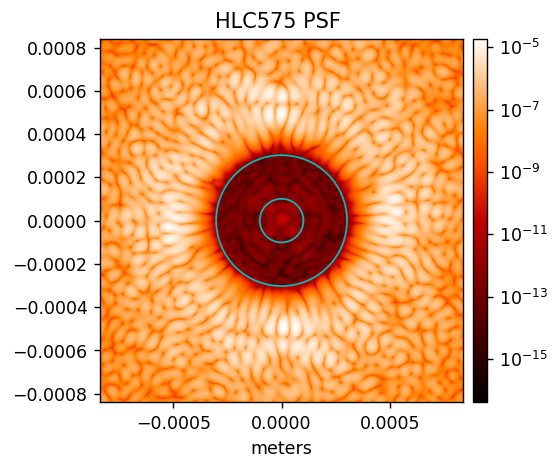}
    \includegraphics[scale=0.45]{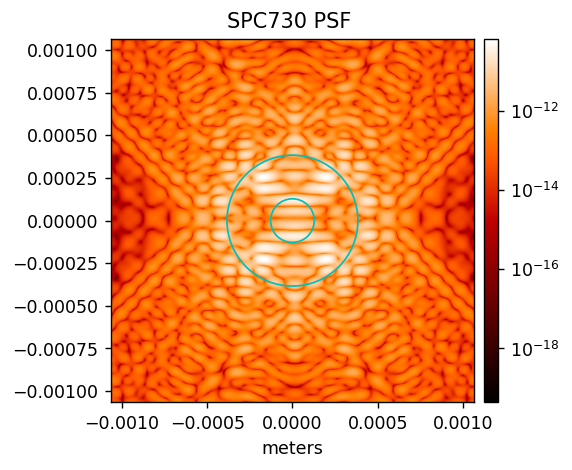}
    \includegraphics[scale=0.45]{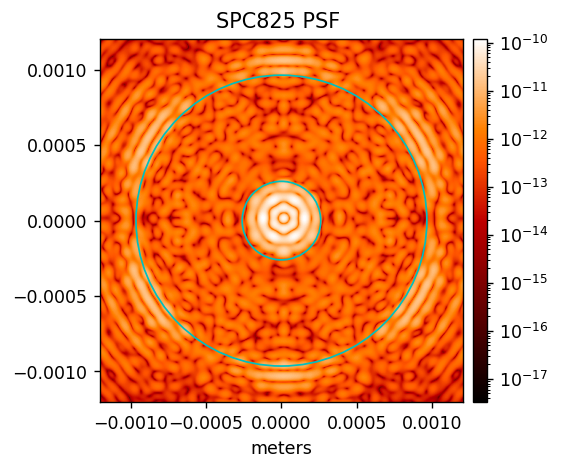}
    \caption{Results from WebbPSF, which differ only in that the resampling routine utilized was from Scipy rather than PROPER.}
    \label{fig:fig}
\end{figure}

Furthermore, as WebbPSF uses POPPY, it can utilize the same options for accelerating the PSF calculations. As a result, the computation times for WebbPSF are equivalent to the computation times reported for the POPPY models.

\section{Conclusions and Future Work}
Overall, the results demonstrated in this paper show that there is practical agreement between the new POPPY models and the previous PROPER models. The primary challenges when creating the POPPY models were implementation differences in how POPPY and PROPER handle pixelscales for different optics and OPDs, the oversampling used for the system and specific optics, and the OPDs caused by the actuator positions provided in the DM maps. Given the sensitivity of the coronagraphic instruments, small difference in pixelscales and interpolation results can cause differing PSF results, which is why these factors must be considered thoroughly when modeling these systems. The discrepancy, namely the on-axis PSFs for the SPC825 mode with OPDs and DMs used, is still being investigated to be understood and corrected for. Other than the PSF results, we have improved the computation times for the two SPC modes tested and while the HLC modes computational performance has been negatively impacted due to the larger array size being propagated for the entire system, improvements are underway in the POPPY configuration to close the remaining performance gaps. 

Another topic that is still being investigated are the generation of polychromatic PSFs. Also, a new implementation of EFC for POPPY Fresnel models is being planned. Finally, updated PROPER models have been released for the final coronagraph Phase C mask designs; thus, updates to the POPPY models will also be tested and released building on the work presented here.

All the code used to generate these results is provided on Zenodo\cite{kian1377-roman-cgi-poppy}. 

\section{Acknowledgments}
The authors acknowledge valuable work done by John Krist, A. J. Eldorado Riggs, and the rest of the JPL and IPAC CGI teams on the creation and distribution of the Roman CGI PROPER models. 

Portions of this work were supported by the Roman Science Investigation team prime award $\#$NNG16PJ24C. Portions of this work were supported by the Arizona Board of Regents Technology Research Initiative Fund (TRIF). This research made use of the High Performance Computing (HPC) resources supported by the University of Arizona (UA) TRIF, UITS, and RDI and maintained by the UA Research Technologies department.

This research made use of community-developed core Python packages, including: Astropy \cite{the_astropy_collaboration_astropy_2013}, Matplotlib \cite{hunter_matplotlib_2007}, SciPy \cite{jones_scipy_2001}, Jupyter, IPython Interactive Computing architecture \cite{perez_ipython_2007,kluyver_jupyter_2016}, FFTW \cite{frigo_fftw_1998} (via pyFFTW v0.12.0), Intel mkl$\_$fft (v1.3.0.post0), and PyOpenCL\footnote{\url{https://github.com/inducer/pyopencl/releases/tag/v2021.2.6}}. Another package used to enable POPPY's PyOpenCL implementation was gpyfft\footnote{\url{https://github.com/geggo/gpyfft/releases/tag/v0.7.0}}.

\newpage
\section{Appendix}
\begin{table}[H]
    \centering
    \caption{List of optics in sequential order for the Roman CGI. These include the Fast-Steering Mirror (FSM), the (FOCM), the Off-Axis Parabolas (OAPs), and fold mirrors. Following the optic is the associated POPPY class used to define the optic as well as its Focal Length (if applicable) and the distance from the previous optic such that POPPY/PROPER can propagate the wavefront to the correct plane.}
    \begin{tabular}{|P{2.5cm}|P{4cm}|P{3cm}|P{3cm}|} \hline
    Optic & POPPY Class & Focal Length [m] & Distance From Previous Optic [m] \\ \hline
    Primary (M1) & QuadraticLens & 2.838 & - \\ \hline 
    Secondary (M2) & QuadraticLens & -0.654 & 2.285 \\ \hline
    Fold 1 & CircularAperture & - & 2.994 \\ \hline
    M3 & QuadraticLens & 0.4302 & 1.681 \\ \hline
    M4 & QuadraticLens & 0.1162 & 0.9435 \\ \hline
    M5 & QuadraticLens & 0.1988 & 0.4291 \\ \hline
    Fold 2 & CircularAperture & - & 0.3511 \\ \hline
    FSM & ScalarTransmission & - & 0.3654 \\ \hline
    OAP 1 & QuadraticLens & 0.5033 & 0.3548 \\ \hline
    FOCM & ScalarTransmission & - & 0.7680 \\ \hline
    OAP2 & QuadraticLens & 0.5791 & 0.3145 \\ \hline
    DM 1 & FITSOpticalElement & - & 0.7758 \\ \hline
    DM 2 & FITSOpticalElement & - & 1.000 \\ \hline
    OAP 3 & QuadraticLens & 1.217 & 0.3948 \\ \hline
    Fold 3 & CircularAperture & - & 0.5053 \\ \hline
    OAP 4 & QuadraticLens & 0.4470 & 1.159 \\ \hline
    Apodizer/SPM & ScalarTransmission or FITSOpticalElement & - & 0.4230 \\ \hline
    OAP 5 & QuadraticLens & 0.5482 & 0.4088 \\ \hline
    FPM & FITSOpticalElement or FITSFPMElement & - & 0.5482 \\ \hline
    OAP 6 & QuadraticLens & 0.5482 & 0.5482 \\ \hline
    Lyotstop & FITSOpticalElement & - & 0.6876 \\ \hline
    OAP 7 & QuadraticLens & 0.7083 & 0.4017 \\ \hline
    Field stop & ScalarTransmission or CircularAperture & - & 0.7083 \\ \hline
    OAP 8 & QuadraticLens & 0.2110 & 0.2110 \\ \hline
    Filter & CircularAperture & - & 0.3684 \\ \hline
    Imaging Lens & QuadraticLens & 0.2960 & 0.1708 \\ \hline
    Fold 4 & CircularAperture & - & 0.2460 \\ \hline
    Image Plane & ScalarTransmission & - & 0.0500 \\ \hline
    \end{tabular}
    \label{tab:optics}
\end{table}

\bibliographystyle{spiebib} 
\bibliography{mybib} 

\begin{thebibliography}{10}

\bibitem{habex}
Gaudi, B.~S., Seager, S., Mennesson, B., Kiessling, A., and Warfield, K.,
  ``{The Habitable Exoplanet Observatory (HabEx)},'' in [{\em UV/Optical/IR
  Space Telescopes and Instruments: Innovative Technologies and Concepts
  IX}{\nolinebreak\hspace{0.1em}]},  Barto, A.~A., Breckinridge, J.~B., and
  Stahl, H.~P., eds.,  {\bf 11115},  125 -- 134, International Society for
  Optics and Photonics, SPIE (2019).

\bibitem{luvoir}
Bolcar, M.~R., ``{The Large UV/Optical/Infrared (LUVOIR) surveyor: engineering
  design and technology overview},'' in [{\em UV/Optical/IR Space Telescopes
  and Instruments: Innovative Technologies and Concepts
  IX}{\nolinebreak\hspace{0.1em}]},  Barto, A.~A., Breckinridge, J.~B., and
  Stahl, H.~P., eds.,  {\bf 11115},  170 -- 183, International Society for
  Optics and Photonics, SPIE (2019).

\bibitem{roman-info}
Kasdin, N.~J., Bailey, V.~P., and Bertrand~Mennesson, e.~a., ``{The Nancy Grace
  Roman Space Telescope Coronagraph Instrument (CGI) technology
  demonstration},'' in [{\em Space Telescopes and Instrumentation 2020:
  Optical, Infrared, and Millimeter Wave}{\nolinebreak\hspace{0.1em}]},
  Lystrup, M., Perrin, M.~D., Batalha, N., Siegler, N., and Tong, E.~C., eds.,
  {\bf 11443},  300 -- 313, International Society for Optics and Photonics,
  SPIE (2020).

\bibitem{krist_proper:_2007}
Krist, J.~E., ``{{PROPER}}: an optical propagation library for {{IDL}},'' in
  [{\em Proc. {{SPIE}}}{\nolinebreak\hspace{0.1em}]},   {\bf 6675},
  66750P--66750P--9 (2007).

\bibitem{krist_overview_2015}
Krist, J., Nemati, B., Zhou, H., and Sidick, E., ``An overview of
  {{WFIRST}}/{{AFTA}} coronagraph optical modeling,'' in [{\em Proc
  SPIE}{\nolinebreak\hspace{0.1em}]},   960505 (Sept. 2015).

\bibitem{krist_wfirst_2017}
Krist, J., Riggs, A.~J., and McGuire, James, e.~a., ``{{WFIRST}} coronagraph
  optical modeling,'' in [{\em Techniques and {{Instrumentation}} for
  {{Detection}} of {{Exoplanets VIII}}}{\nolinebreak\hspace{0.1em}]},   {\bf
  10400},  1040004, {International Society for Optics and Photonics} (Sept.
  2017).

\bibitem{krist_wfirst_2018}
Krist, J., Effinger, R., and Kern, Brian, e.~a., ``{{WFIRST}} coronagraph
  flight performance modeling,'' in [{\em Space {{Telescopes}} and
  {{Instrumentation}} 2018: {{Optical}}, {{Infrared}}, and {{Millimeter
  Wave}}}{\nolinebreak\hspace{0.1em}]},   {\bf 10698},  106982K, {International
  Society for Optics and Photonics} (July 2018).

\bibitem{falco}
Riggs, A. J.~E., Ruane, G., Sidick, E., Coker, C., Kern, B.~D., and Shaklan,
  S.~B., ``{Fast linearized coronagraph optimizer (FALCO) I: a software toolbox
  for rapid coronagraphic design and wavefront correction},'' in [{\em Space
  Telescopes and Instrumentation 2018: Optical, Infrared, and Millimeter
  Wave}{\nolinebreak\hspace{0.1em}]},  Lystrup, M., MacEwen, H.~A., Fazio,
  G.~G., Batalha, N., Siegler, N., and Tong, E.~C., eds.,  {\bf 10698},  878 --
  888, International Society for Optics and Photonics, SPIE (2018).

\bibitem{basic-wf-theory}
Wyant, J. and Creath, K., ``Basic wavefront aberration theory for optical
  metrology,'' {\em Appl Optics Optical Eng}~{\bf 11} (01 1992).

\bibitem{Soummer:07}
Soummer, R., Pueyo, L., Sivaramakrishnan, A., and Vanderbei, R.~J., ``Fast
  computation of lyot-style coronagraph propagation,'' {\em Opt. Express}~{\bf
  15},  15935--15951 (Nov 2007).

\bibitem{webbpsf}
Perrin, M.~D., Soummer, R., Elliott, E.~M., Lallo, M.~D., and Sivaramakrishnan,
  A., ``{Simulating point spread functions for the James Webb Space Telescope
  with WebbPSF},'' in [{\em Space Telescopes and Instrumentation 2012: Optical,
  Infrared, and Millimeter Wave}{\nolinebreak\hspace{0.1em}]},  Clampin, M.~C.,
  Fazio, G.~G., MacEwen, H.~A., and Jr., J. M.~O., eds.,  {\bf 8442},  1193 --
  1203, International Society for Optics and Photonics, SPIE (2012).

\bibitem{kian1377-roman-cgi-poppy}
Milani, K., ``{kian1377/Roman-CGI-POPPY: Roman-CGI-POPPY Initial Release with
  Basic Functionality},'' (July 2021).

\bibitem{the_astropy_collaboration_astropy_2013}
{The Astropy Collaboration}, e.~a., ``Astropy: {A} community {Python} package
  for astronomy,'' {\em Astronomy \& Astrophysics}~{\bf 558},  A33 (Oct. 2013).

\bibitem{hunter_matplotlib_2007}
Hunter, J.~D., ``Matplotlib: {A} {2D} graphics environment,'' {\em Computing In
  Science \& Engineering}~{\bf 9}(3),  90--95 (2007).

\bibitem{jones_scipy_2001}
Jones, E., Oliphant, T., and Peterson, P., ``{SciPy}: {Open} source scientific
  tools for {Python},'' {\em http://www. scipy. org/}  (2001).

\bibitem{perez_ipython_2007}
P{\'e}rez, F. and Granger, B., ``{IPython}: {A} {System} for {Interactive}
  {Scientific} {Computing},'' {\em Computing in Science Engineering}~{\bf 9},
  21--29 (May 2007).

\bibitem{kluyver_jupyter_2016}
Kluyver, T., Ragan-Kelley, B., and P{\'e}rez, Fernando, e.~a., ``Jupyter
  {Notebooks}-a publishing format for reproducible computational workflows.,''
  in [{\em Positioning and {Power} in {Academic} {Publishing}: {Players},
  {Agents} and {Agendas}}{\nolinebreak\hspace{0.1em}]},   87--90 (2016).

\bibitem{frigo_fftw_1998}
Frigo, M. and Johnson, S., ``{{FFTW}}: an adaptive software architecture for
  the {{FFT}},'' in [{\em Proceedings of the 1998 {{IEEE International
  Conference}} on {{Acoustics}}, {{Speech}} and {{Signal Processing}},
  {{ICASSP}} '98 ({{Cat}}. {{No}}.{{98CH36181}})}{\nolinebreak\hspace{0.1em}]},
    {\bf 3},  1381--1384 vol.3 (May 1998).

\bibitem{frigo_design_2005}
Frigo, M. and Johnson, S.~G., ``The {{Design}} and {{Implementation}} of
  {{FFTW3}},'' {\em Proceedings of the IEEE}~{\bf 93},  216--231 (Feb. 2005).

\end{thebibliography}
\nocite{*}

\end{document}